\documentclass[lettersize,journal]{IEEEtran}
\IEEEoverridecommandlockouts
\usepackage{cite}
\usepackage{amsmath,amsfonts}
\usepackage{algorithmic}
\usepackage{graphicx}
\usepackage{textcomp}
\usepackage{xcolor}
\usepackage[pscoord]{eso-pic}
\usepackage{lipsum}
\usepackage{mathtools}
\usepackage{amsmath,amssymb,amsfonts}
\usepackage{textcomp}
\usepackage{xcolor} 
\usepackage{booktabs} 
\usepackage{enumitem}
\usepackage{algorithmic,algorithm,amsmath}
\usepackage{graphicx,multirow}
\usepackage{epstopdf}
\usepackage{diagbox}
\usepackage{afterpage}
\usepackage{svg}
\usepackage{subfig}		
\usepackage{xcolor}
\usepackage{pgfplots}
\pgfplotsset{compat=1.15}
\usepackage{verbatim} 
\usepackage{multicol}
\usepackage{hyphenat}
\usepackage{lipsum}
\usepackage{balance}
\usepackage{enumitem}
\usepackage{etoolbox}
\usepackage{euscript}
\usepackage{eufrak}
\usepackage{mathrsfs}
\usepackage{textcase}
\usepackage{bm}
\usepackage[english]{babel}
\usepackage{blindtext}
\usepackage{textcase}
\usepackage{color}
\usepackage{afterpage}
\usepackage{tikz}

\usepackage[tablename=TABLE, tableposition=top]{caption}
\usepackage{pifont}
\usepackage{makecell}
\usepackage[T1]{fontenc}
\usepackage[hidelinks]{hyperref}
\usepackage[acronym]{glossaries}
\usepackage[nodayofweek]{datetime}
\usepackage[compact]{titlesec}

\usepackage{titlefoot}
\DeclareMathAlphabet{\mathpzc}{OT1}{pzc}{m}{it}

\usepackage[all]{background}
\SetBgContents{\textit{This article is under peer-review} }
\SetBgScale{2}
\SetBgColor{black}
\SetBgAngle{0}
\SetBgOpacity{0.7}
\SetBgPosition{current page.north east}
\SetBgVshift{-0.5cm}
\SetBgHshift{-4.75cm}
\addto\captionsngerman{}
\addto\captionsngerman{}
\addto\captionsngerman{}
\addto\captionsngerman{}
\addto\extrasenglish{%

  \renewcommand{\figurename}{Fig.}
}
\newcommand{\cmark}{\ding{51}}%
\newcommand{\xmark}{\ding{55}}%

\usepackage{mdwlist}

\makeglossaries
\newacronym{axc}{AxC}{\text{A}pproximate \text{C}omputing}
\newacronym{axo}{AxO}{\text{A}pproximate \text{O}perator}
\newacronym{ai}{AI}{\text{A}rtificial \text{I}ntelligence}
\newacronym{cpd}{CPD}{\text{c}ritical \text{p}ath \text{d}elay}
\newacronym{pdp}{PDP}{\text{P}ower-\text{D}elay \text{P}roduct}
\newacronym{ilp}{ILP}{\text{I}nstruction \text{L}evel \text{P}arallelism}
\newacronym{vlsi}{VLSI}{\text{V}ery \text{L}arge \text{S}cale \text{I}ntegration}
\newacronym{mttf}{MTTF}{\text{M}ean \text{T}ime \text{T}o \text{F}ailure}
\newacronym{mttc}{MTTC}{\text{M}ean \text{T}ime \text{T}o \text{C}rash}
\newacronym{noc}{NoC}{\text{N}etwork-on-\text{C}hip}
\newacronym{pr}{PR}{\text{P}artially \text{R}econfigurable}
\newacronym{cots}{COTS}{commercial-off-the-shelf}
\newacronym{nre}{NRE}{\text{N}on \text{R}ecurring \text{E}ngineering}
\newacronym{rtl}{RTL}{\text{R}egister \text{T}ransfer \text{L}evel}
\newacronym{vcu}{VCU}{\text{V}ideo \text{C}odec \text{U}nit}
\newacronym{apu}{APU}{\text{A}pplication \text{P}rocessing \text{U}nit}
\newacronym{gpu}{GPU}{\text{G}raphics \text{P}rocessing \text{U}nit}
\newacronym{rpu}{RPU}{\text{R}eal-time \text{P}rocessing \text{U}nit}
\newacronym{gps}{GPS}{\text{G}lobal \text{P}ositioning \text{S}ystem}
\newacronym{ml}{ML}{\text{M}achine \text{L}earning}
\newacronym{iot}{IoTs}{\text{I}nternet of \text{T}hings}
\newacronym{ic}{ICs}{\text{i}ntegrated \text{c}ircuits}
\newacronym{esl}{ESL} {\text{E}lectronic \text{S}ystem \text{L}evel}
\newacronym{eda}{EDA} {\text{E}lectronic \text{D}esign \text{A}utomation}
\newacronym{clr}{CLR} {\text{C}ross\hyp\text{l}ayer \text{R}eliability}
\newacronym{qos}{QoS} {\text{Q}uality of \text{S}ervice}
\newacronym{hmpsoc}{HMPSoC} {\text{H}eterogeneous \text{M}ulti-\text{P}rocessor \text{S}ystem\hyp on\hyp\text{C}hip}
\newacronym{mpsoc}{MPSoC} {\text{M}ulti-\text{P}rocessor \text{S}ystem\hyp on\hyp\text{C}hip}
\newacronym{soc}{SoC} {\text{S}ystem\hyp \text{o}n\hyp\text{C}hip}
\newacronym{fpga}{FPGA} {\text{F}ield \text{P}rogrammable \text{G}ate \text{A}rray}
\newacronym{dpr}{DPR} {\text{D}ynamic \text{P}artial \text{R}econfiguration}
\newacronym{prr}{PRR} {\text{P}artially \text{R}econfigurable \text{R}egion}
\newacronym{prm}{PRM} {\text{P}artially \text{R}econfigurable \text{M}odule}
\newacronym{pe}{PE} {\text{P}rocessing \text{E}lement}
\newacronym{dse}{DSE} {\text{D}esign \text{S}pace \text{E}xploration}
\newacronym{ga}{GA} {\text{G}enetic \text{A}lgorithms}
\newacronym{bti}{BTI} {\text{B}ias \text{T}emperature \text{I}nstability}
\newacronym{nbti}{NBTI} {\text{N}egative \text{B}ias \text{T}emperature \text{I}nstability}
\newacronym{pbti}{PBTI} {\text{P}ositive \text{B}ias \text{T}emperature \text{I}nstability}
\newacronym{em}{EM} {\text{E}lectro\text{m}igration}
\newacronym{gob}{GOB} {\text{G}ate \text{O}xide \text{B}reakdown}
\newacronym{hci}{HCI} {\text{H}ot \text{C}arrier \text{I}njection}
\newacronym{tddb}{TDDB}{\text{T}ime \text{D}ependent \text{D}ielectric \text{B}reakdown}
\newacronym{seu}{SEU} {\text{S}ingle \text{E}vent \text{U}pset}
\newacronym{ser}{SER} {\text{S}oft \text{E}rror \text{R}ate}
\newacronym{gdb}{GDB} {\text{G}ate \text{D}ielectric \text{B}reakdown}
\newacronym{tmr}{TMR} {\text{T}riple \text{M}odular \text{R}edundancy}
\newacronym{dmr}{DMR} {\text{D}ual \text{M}odular \text{R}edundancy}
\newacronym{ecc}{ECC}{\text{E}rror \text{C}hecking and \text{C}orrecting}
\newacronym{sram}{SRAM}{\text{S}tatic \text{R}andom \text{A}ccess \text{M}emory}
\newacronym{dram}{DRAM}{\text{D}ynamic \text{R}andom \text{A}ccess \text{M}emory}
\newacronym{llc}{LLC}{\text{L}ast \text{L}evel \text{C}ache}
\newacronym{l1}{L1}{\text{L}evel \text{1}}
\newacronym{dimm}{DIMM}{\text{D}ual \text{i}n-line-\text{M}emory \text{M}odule}
\newacronym{snc}{SNC}{\text{S}ingle-\text{N}ibble-error-\text{C}orrecting}
\newacronym{dnd}{DND}{\text{D}ouble-\text{N}ibble-error-\text{D}etecting}
\newacronym{sec}{SEC}{\text{S}ingle-bit-\text{E}rror-\text{C}orrecting}
\newacronym{ded}{DED}{\text{D}ouble-bit-\text{E}rror-\text{D}etecting}
\newacronym{dec}{DEC}{\text{D}ouble-bit-\text{E}rror-\text{C}orrecting}
\newacronym{ted}{TED}{\text{T}riple-bit-\text{E}rror-\text{D}etecting}
\newacronym{ivi}{IVI}{\text{I}nstruction \text{V}ulnerability \text{I}ndex}
\newacronym{fvi}{FVI}{\text{F}unction \text{V}ulnerability \text{I}ndex}
\newacronym{sed}{SED}{\text{S}obel \text{E}dge \text{D}etection}
\newacronym{cnn}{CNN}{\text{C}onvolutional \text{N}eural \text{N}etworks}
\newacronym{dnn}{DNN}{\text{D}eep \text{N}eural \text{N}etworks}
\newacronym{os}{OS}{\text{O}perating \text{S}ystem}
\newacronym{avf}{AVF}{\text{A}rchitectural \text{V}ulnerability \text{F}actor}
\newacronym{milp}{MILP}{\text{M}ixed \text{I}nteger \text{L}inear \text{P}rogramming}
\newacronym{sofr}{SOFR}{\text{S}um-\text{o}f-\text{F}ailure \text{R}ate}
\newacronym{clb}{CLB}{\text{C}onfigurable \text{L}ogic \text{B}locks}
\newacronym{bram}{BRAM}{\text{B}lock \text{RAM}}
\newacronym{dsps}{DSPs}{\text{D}igital \text{S}ignal \text{P}rocessing blocks}
\newacronym{mcts}{MCTS}{\text{M}onte \text{C}arlo \text{T}ree \text{S}earch}
\newacronym{ttp}{TTP} {\text{T}ree \text{T}raversal \text{P}roblem}
\newacronym{fir}{FIR} {\text{F}inite \text{I}mpluse \text{R}esponse}
\newacronym{mtbf}{MTBF}{Mean Time between Failures}
\newacronym{ura}{\textit{uRA}}{User-modulated Run-time Adaptation}
\newacronym{aura}{\textit{AuRA}}{Agent-based User-modulated Run-time Adaptation}
\newacronym{moea}{MOEA}{\text{M}ulti-\text{O}bjective \text{E}volutionary \text{A}lgorithms}
\newacronym{dvfs}{DVFS}{\text{D}ynamic \text{V}oltage and \text{F}requncy \text{S}caling}
 \newacronym{icap}{ICAP}{\text{I}nternal \text{C}onfiguration \text{A}ccess \text{P}ort}
\newacronym{rl}{RL}{\text{R}einforcement \text{L}earning}
\newacronym{pvt}{PVT}{\text{P}rocess, \text{V}oltage, and \text{T}emperature}
\newacronym{nhpp}{NHPP}{\text{N}on-\text{H}omogeneous \text{P}oisson \text{P}rocess}
\newacronym{fit}{FIT}{\text{F}ailures \text{I}n \text{T}ime}
\newacronym{mosfet}{MOSFET}{\text{M}etal \text{O}xide \text{S}emiconductor \text{F}ield \text{E}ffect \text{T}ransistor }
\newacronym{nmos}{NMOS}{\text{N}egative channel \text{M}etal \text{O}xide \text{S}emiconductor}
\newacronym{pmos}{PMOS}{\text{P}ositive channel \text{M}etal \text{O}xide \text{S}emiconductor}
\newacronym{osi}{OSI}{\text{O}pen \text{S}ystems \text{I}nterconnection}
\newacronym{bist}{BIST}{\text{B}uilt-\text{i}n \text{S}elf \text{T}est}
\newacronym{nlp}{NLP}{\text{N}atural \text{L}anguage \text{P}rocessing}
\newacronym{dof}{DoF}{\text{D}egree \text{o}f \text{F}reedom}
\newacronym{lut}{LUT}{Look-Up Table}
\newacronym{ppa}{PPA}{Power-Performance-Area}
\newacronym{behav}{BEHAV}{behavioral accuracy}
\newacronym{mse}{MSE}{Mean Squared Error}
\newacronym{mlp}{MLP}{Multi-Layer Perceptron}
\newacronym{l2}{L2}{Squared Error}
\newacronym{ppf}{PPF}{Pseudo Pareto-front}
\newacronym{vpf}{VPF}{Validated Pareto-front}
\newacronym{rmse}{RMSE}{Root Mean Squared Error}
\newacronym{mac}{MAC}{Multiply-Accumulate}
\newacronym{asic}{ASIC}{Application-specific Integrated Circuit}
\newacronym{gan}{GAN}{Generative Adversarial Network}
\newacronym{ann}{ANN}{Artificial Neural Network}
\newacronym{lpf}{LPF}{Low-pass Filter}
\newacronym{ecg}{ECG}{Electrocardiogram}
\newacronym{shap}{SHAP}{SHapley Additive exPlanations}
\newacronym{bfs}{BFS}{Breadth First Search}
\newacronym{dfs}{DFS}{Depth First Search}
\newacronym{csp}{CSP}{Constraint Satisfaction Problem}
\newacronym{cgp}{CGP}{Cartesian Genetic Programming}
\newacronym{cc}{CC}{Carry-propagation-Chain}
\newacronym{conss}{ConSS}{Configuration Supersampling}
\begin{document}
\newcommand{\doctitle}{
    \textit{AxOCS}: Scaling FPGA-based \underline{A}ppro\underline{x}imate \underline{O}perators using \underline{C}onfiguration \underline{S}upersampling 
}
\newcommand{\slidesloc}{https://www.dropbox.com/s/s29xy7xfw5ul6cw/outline.pptx?dl=0}
\newcommand{\titleName}{\textit{AxOCS}}
\newcommand{\lws}{LUT-wise~significance~}

\newcommand\mylistkeywords{
AI-based Design Space Exploration,
Approximate Computing, 
Arithmetic Operator Design, 
Circuit Synthesis,
}

\newcommand{\add}[1]{\textcolor{red}{#1}}
\newcommand{\rephrase}[1]{\textcolor{black}{#1}}
\newcommand{\siva}[1]{\textcolor{red}{#1}}
\newcommand{\rev}[1]{\textcolor{black}{#1}}
\newcommand{\salim}[1]{\textcolor{black}{#1}}
\newcommand{\aspdac}[1]{\textcolor{black}{#1}}
\newcommand{\todo}[1]{\textcolor{red}{#1}}
	
\title{
        \doctitle  
}

\IEEEtitleabstractindextext{
\begin{abstract}
\rev{
The rising usage of AI/ML-based processing across application domains has exacerbated the need for low-cost ML implementation, specifically for resource-constrained embedded systems. \salim{To this end,} approximate computing, an approach that explores the power, performance, area (PPA), and behavioral accuracy (BEHAV) trade-offs, has emerged as a possible solution for implementing embedded machine learning. Due to the predominance of MAC operations in ML, designing platform-specific approximate arithmetic operators forms one of the major research problems in approximate computing. Recently there has been \salim{a} rising usage of AI/ML-based design space exploration \salim{techniques} for implementing approximate operators. However, \salim{most of these} approaches are limited to using ML-based surrogate functions for predicting the PPA and BEHAV impact of a set of related design decisions. While this approach leverages the regression capabilities of ML methods, it does not exploit the more advanced approaches in ML. To this end, we propose \titleName, a methodology for designing approximate arithmetic operators through ML-based supersampling. Specifically, we present a method to leverage the correlation of PPA and BEHAV metrics across operators of varying bit-widths for generating larger bit-width operators. 
\salim{The proposed approach involves traversing the relatively smaller design space of smaller bit-width operators and employing its associated \textit{Design-PPA-BEHAV} relationship to generate initial solutions for metaheuristics-based optimization for larger operators.}
\salim{The experimental evaluation of \titleName~for FPGA-optimized approximate operators shows that the proposed approach significantly improves the quality---resulting hypervolume for multi-objective optimization---of  $8\times8$ signed approximate multipliers.}
}

\end{abstract}
}

\author{Siva Satyendra Sahoo, Salim Ullah, Soumyo Bhattacharjee and Akash Kumar,
\IEEEmembership{Senior Member, IEEE}
}

\newcounter{tblEqCounter} 
\maketitle 

\unmarkedfntext{© 2023 IEEE. Personal use of this material is permitted. Permission
from IEEE must be obtained for all other uses, in any current or future
media, including reprinting/republishing this material for advertising or
promotional purposes, creating new collective works, for resale or
redistribution to servers or lists, or reuse of any copyrighted
component of this work in other works.}
\IEEEdisplaynontitleabstractindextext


\section{Introduction}
\label{sec:intro}
\rev{
\salim{\gls{axc} allows for disproportionate gains in the \gls{ppa} of an application's implementation by allowing some level of reduction in its \gls{behav}. \gls{axc} usually refers to a host of methods that can be implemented across multiple layers of the computation stack and includes techniques such as loop skipping, input scaling, and truncation~\cite{10.1145/2893356}. }
All these methods involve the deliberate introduction of some form of error/inaccuracy in processing while staying within the application's error tolerance constraints 
to improve the system's \gls{ppa} metrics. For the hardware layer
\salim{of the computation stack}, designing approximate arithmetic operators forms one of the more effective methods of implementing approximations in embedded systems\salim{~\cite{chippa2014scalable,9586260}.}
}
\rev{
Arithmetic operations form an essential component of \salim{various applications, especially} \gls{ml} inference, with \gls{mac} representing the primitive operation of most inference processing. Hence, low-cost implementation of computer arithmetic, primarily multipliers and adders, is an active area of research for embedded \gls{ml}. Designing \glspl{axo} usually involves avoiding the more costly binary operations in the operators' algorithm, resulting in implementations with varying levels of \gls{ppa} and \gls{behav} trade-offs. 
\salim{Although similar to precision scaling regarding design goals, \glspl{axo} provide a far larger scope for a finer granularity of optimization~\cite{10.1145/2893356,9344673}.}
}

\begin{figure}[t]
    \centering
  \subfloat[Absolute metrics]{
    \scalebox{1.0}{\includegraphics[width=0.34 \columnwidth]{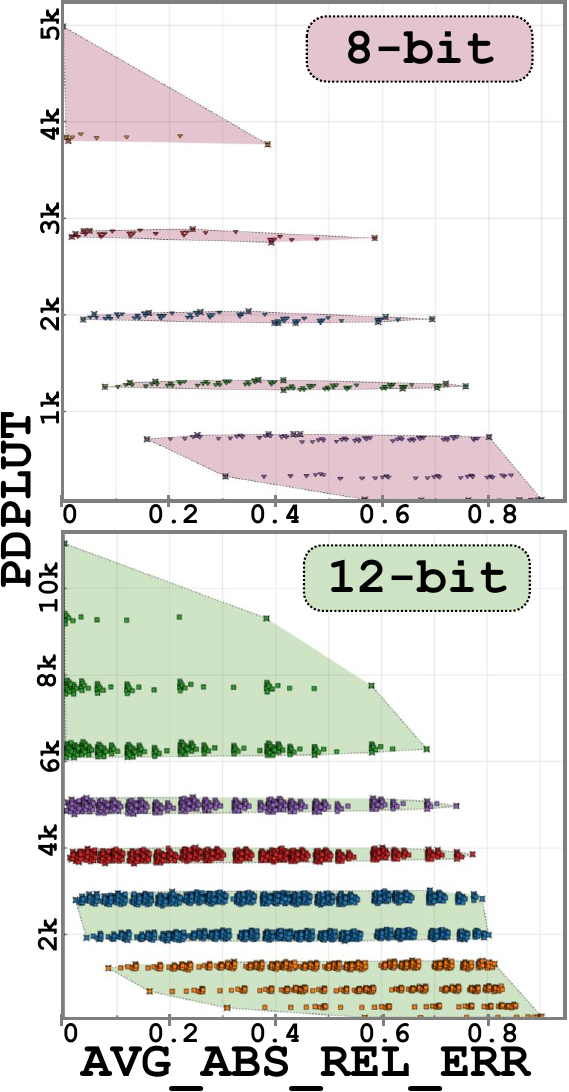}}
       }
  \subfloat[Scaled metrics]{%
        \scalebox{1.0}{\includegraphics[width=0.66 \columnwidth]{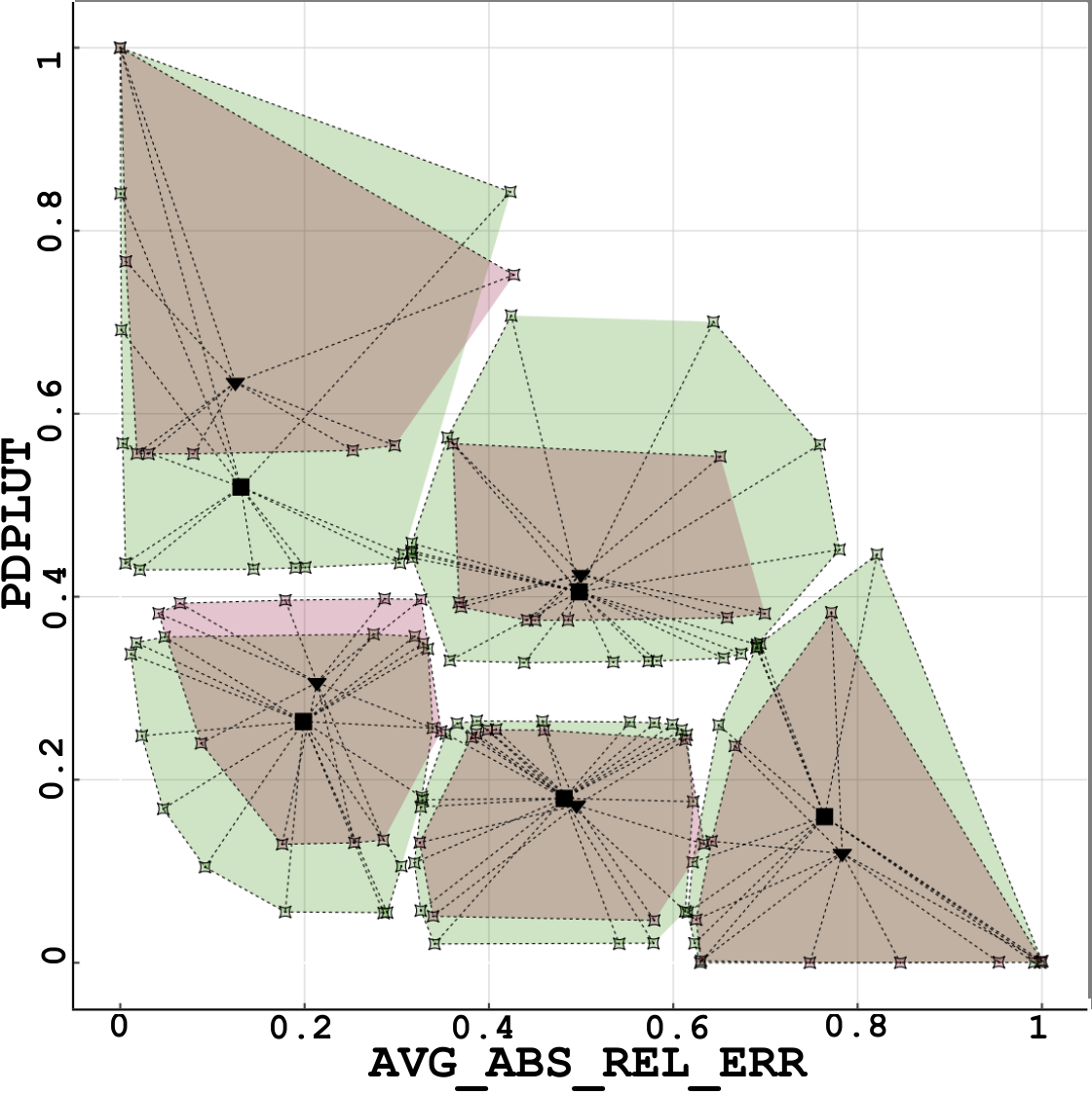}}
        }
\caption{k-means clustering of designs points representing approximate implementations of 8-bit and 12-bit unsigned adders}
  \label{fig:mot} 
\end{figure}

    \begin{figure*}[t] 
        \centering
            \scalebox{1.0}{\includegraphics[width=1.75 \columnwidth]{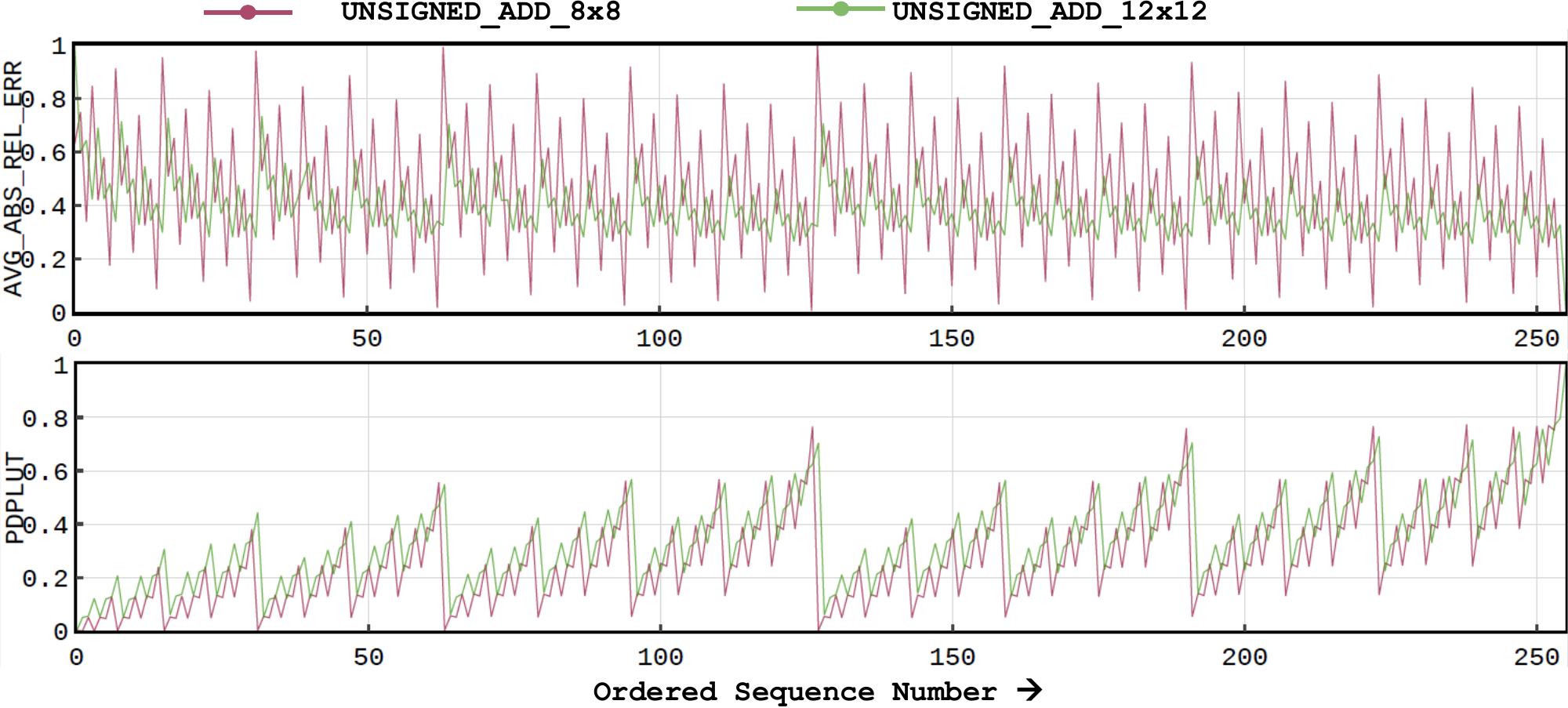}}
    \caption{Variation of scaled PDPLUT and AVG\_ABS\_REL\_ERR with UINT-encoded configuration for 8-bit and 12-bit unsigned approximate adders. The ordered (based on UINT configuration) sequence of metrics for 12-bit designs are sub-sampled to get a similar length sequence for both operators.}
      \label{fig:mot_2} 
    \end{figure*}

\rev{
However, this larger scope of optimization comes at the cost of a large design space, further exacerbated by the consideration of the hardware platform during \gls{dse}. Early works on \gls{dse} for \glspl{axo} had focused on \gls{asic}-based hardware platforms\salim{~\cite{hashemi2015drum, 9233379}}. However, with the rising diversity of application domains implementing \gls{ml} and the ever-evolving nature of \gls{ml} algorithms, \glspl{fpga} find increasing prevalence as the hardware platform across different scales of computing--from TinyML~\cite{tinyml} to cloud computing~\cite{8977886}. Consequently, more recent works have also focused on designing \gls{fpga}-based \glspl{axo}. Corresponding approaches range from implementing \gls{asic}-optimized logic on \glspl{fpga} to optimizing specifically for \glspl{fpga}' \gls{lut} and \gls{cc}-based architecture\salim{~\cite{9218533, 10.1145/3195970.3196115}}. Recently multiple related works have reported the use of AI/ML-based methods in the \gls{dse} for \gls{fpga}-based \glspl{axo}\salim{~\cite{mrazek2019autoax, ullah2022appaxo}}. 
However, none of the related works \salim{analyze and} exploit any correlation across operators of different bit-widths.
}



\rev{
For instance, \autoref{fig:mot} shows the results of k-means clustering of different \gls{axo} implementations of unsigned 8-bit and 12-bit adders. The \glspl{axo} were obtained using the operator model presented in~\cite{ullah2022appaxo}. It involves the selective removal of a subset of \glspl{lut} used in the accurate implementation and hence results in 256 (=2\textsuperscript{8}) and 4096 (=2\textsuperscript{12}) designs for the 8-bit and 12-bit unsigned adders, respectively. Each \salim{approximate design} is represented by its PDPLUT\footnote{PDPLUT = Power $\times$ CPD $\times$ LUT usage} and AVG\_ABS\_REL\_ERR\footnote{AVG\_ABS\_REL\_ERR: Average absolute relative error} as the representative \gls{ppa} and \gls{behav} metrics, respectively. \autoref{fig:mot}(a) plots all the possible design points in terms of their absolute metrics, along with the 5 clusters, obtained from the elbow method-based selection of k-means clustering. \autoref{fig:mot}(b) shows the clusters, in terms of the centroid and the convex hull of design points in each cluster, obtained by k-means clustering with the \textit{min-max scaled} metrics of the 8-bit and 12-bit unsigned adder \glspl{axo}. As evident from the figure, the centroids of the clusters are in the vicinity of each other for both the bit-width \glspl{axo}. In addition, \autoref{fig:mot}(b) shows similar clusters and the overall spread of the design points for both bit-width operators.
}



\rev{
\autoref{fig:mot} demonstrates some possible underlying patterns in the \gls{ppa} and \gls{behav} metrics' distribution across operators of different bit widths. However, it does not provide any information regarding the correlation of the approximate configuration, representing the combination of \glspl{lut} that are selected for removal/usage\footnote{Using 1/0 to represent the usage/removal of each LUT respectively}, and the corresponding design metrics. 
}
\salim{\autoref{fig:mot_2} shows the variation of scaled PDPLUT and AVG\_ABS\_REL\_ERR for 8- and 12-bit unsigned adders. For the 8-bit adder, we have plotted the individual scaled \gls{ppa} and \gls{behav} metrics for all 256 configurations. However, for the 12-bit adder (having 4096 configurations), we show the mean value of the scaled \gls{ppa} and \gls{behav} metrics in non-overlapping consecutive windows of 16 configurations. This sampling technique for 12-bit adders provides 256 data points and helps analyze the trends and variation of \gls{ppa} and \gls{behav} metrics of arithmetic operators of different bit-widths. As evident from \autoref{fig:mot_2}, the \textit{Configuration-\gls{ppa}-\gls{behav}} values demonstrate similar patterns for both bit-width operators. However, the related state-of-the-art works do not employ such statistical analysis of the characterization data across different bit-width operators in the \gls{dse} of \glspl{axo}. To this end, we propose \titleName, a methodology that utilizes modern \gls{ml} techniques in the \gls{dse} of \gls{fpga}-based \glspl{axo}. \\ }
\rev{
\textbf{Contributions:}
\begin{enumerate}
    \item We present a statistical analysis of the characterization data of \gls{fpga}-based \glspl{axo} across different bit-widths. Specifically, we present methods for investigating the patterns and relationships between the approximate configuration and the PPA and BEHAV metrics of approximate implementations. 
    \item We present an ML-based supersampling of approximate design configurations. \salim{In particular}, we use the presented statistical analysis to generate larger bit-width operators from smaller bit-width approximate operators' characterization data. 
    \item We present an augmented metaheuristics-based optimization method for the corresponding \gls{dse} problem. Specifically, we use results from the proposed supersampling to direct a \gls{ga}-based multi-objective optimization. 
\end{enumerate}
}

\rev{
The rest of the article is organized as follows.
\autoref{sec:bckRel} presents a brief overview of the requisite background and related works. 
The operator model used in the analysis is presented in~\autoref{sec:op_model}. 
\autoref{sec:propDSE} presents the various components and methods of the proposed \titleName~ methodology. 
The analysis of the 
experimental evaluation of the proposed methods is presented in~\autoref{sec:expRes}. 
Finally, in~\autoref{sec:conc}, we conclude the article with a summary and a brief discussion of the scope for related future research.
}

   %
\section{Background and Related Works}
\label{sec:bckRel}
\subsection{Approximate Computing}
\label{subsec:AxC}
\rev{
\acrlong{axc} has emerged as a potential solution for the ever-increasing computational and memory demands of modern applications.
Most of these applications are characterized by their inherent resilience to inaccuracies in data representation and related intermediate computations. 
The inherent error resilience of these applications enables them to produce multiple feasible answers instead of one golden answer. 
\gls{axc} exploits this error resilience by providing designs with varying trade-offs between output 
accuracy and the system implementation's performance gains. 
Approximation techniques covering multiple layers of the computation stack have been the focus of a host of recent related works~\cite{10.1145/2893356, yin2017minimizing, miguel2015doppelganger, chippa2014scalable, venkataramani2015computing, Ullah2023}.
}

\rev{
\salim{Among the various layers of the computation stack, approximation techniques related to the} architecture and circuit layers have acquired the most significant attention for implementing compute-intensive \salim{error-tolerant} applications on resource-constrained embedded systems. 
For the architecture level, reduced precision computer arithmetic and storage have emerged as the most widely utilized techniques~\cite{wang2019bfloat16,yin2017minimizing}. 
Similarly, at the circuit layer, using inaccurate computational units is one of the more effective techniques. 
\salim{As \gls{mac} is one of the primary operations in various error-resilient applications, such as \gls{ml} inference, }
most related works have focused on the approximate implementation of adders and multipliers~\cite{6691096, 8342140, 9233379, 5771062, hashemi2015drum, 10.1145/3195970.3196115, 9344673, 9072581,7827657}. 
Most of these approaches involve either truncating parts of computation or utilizing inaccurate computations to introduce deliberate approximations for performance gains. 
For example, implementing multiple sub-adders to truncate the long carry-propagation-chain in a larger adder is used in~\cite{6691096}. 
Similarly, utilizing different carry and sum prediction techniques to implement a set of \gls{fpga}-optimized approximate adders is presented in~\cite{8342140}.
}


\rev{
Given the high implementation cost of multiplication operations, most related works have proposed various \gls{asic}- and \gls{fpga}-based approximate multiplier architectures. 
For instance, the authors of~\cite{5771062} use truncation methods to produce $M$-bit output for an $\text{M}\times\text{M}$ multiplier. 
Similarly, the authors of~\cite{9233379} have utilized \gls{cgp} to present libraries of \gls{asic}-optimized approximate adders and multipliers. 
For this purpose, the accurate circuits are represented using a string of integers, and a worst-case error-based objective function is used to generate various approximate versions of a circuit.
Similarly, some works, such as~\cite{hashemi2015drum}, truncate the input operands to employ a smaller multiplier to implement a larger multiplier. 
\salim{Other works, such as~\cite{7827657}, 
\salim{employs}
functional approximation to implement 
\gls{asic}-optimized 
inaccurate 
$2\times2$ 
multipliers, which are then 
used to implement larger multipliers.
}}

\subsection{DSE for FPGA-based Approximate Operators}
\label{sub:ai_ml_dse}
\begin{table*}[t]
\centering
\caption{Comparing related works 
}
\label{table:rel}
\def\arraystretch{1.2}
\resizebox{0.7 \textwidth}{!}{
\begin{tabular}{@{}ccccccccc@{}}
\toprule
Related Works & \cite{9233379} & \cite{9072581} & \cite{mrazek2019autoax} & \cite{9218533} & \cite{ullah2022appaxo} &  \cite{10.1145/3566097.3567891} & \cite{10.1145/3195970.3196115,9344673,8465781}& \titleName \\ \midrule
LUT-level Optimization & \xmark & \cmark & \xmark & \xmark & \cmark & \cmark & \cmark & \cmark \\ \midrule
Automated Search & \cmark & \xmark & \cmark & \xmark & \cmark & \cmark & \xmark & \cmark \\ \midrule
ML-based Estimation & \xmark & \xmark & \cmark & \cmark & \cmark & \cmark & \xmark & \cmark \\ \midrule
Iterative Search & \cmark & \cmark & \cmark & \cmark & \cmark & \xmark & \xmark & \cmark \\ \midrule
Operator Scaling & \xmark & \xmark & \cmark & \cmark & \cmark & \xmark & \cmark & \cmark \\ \midrule
Directed Search & \xmark & \xmark & \xmark & \xmark & \xmark & \xmark & \xmark & \cmark \\ \bottomrule
\end{tabular}
}
\end{table*}


\rev{
The related works discussed earlier implement approximate operators primarily for \gls{asic}-based hardware platforms.
The works presented in~\cite{9344673, 10.1145/3195970.3196115, 9072581} have utilized the \glspl{lut} and \glspl{cc} structures of Xilinx \glspl{fpga} to propose various approximate multiplier architectures. 
The authors \salim{of~\cite{9344673, 8465781} have presented single designs of} $4\times4$ approximate multiplier to implement higher-order multipliers. 
Similarly, in~\cite{10.1145/3195970.3196115}, the authors utilize three different designs of \gls{fpga}-optimized approximate $4\times4$ multipliers to implement a higher-order approximate multiplier library. 
However, this work does not provide an intelligent/automated \gls{dse} mechanism for identifying approximate implementations that provide better accuracy-performance trade-offs. 
The authors of~\cite{9072581} have utilized radix-4 Booth's algorithm to present a single approximate signed multiplier implementation. 
Their proposed implementation employs both functional approximation and truncation of product bits to reduce the total resource utilization.
However, the removal of \glspl{lut} is based on a manual search of the most power-consuming components of the accurate implementation.
}

\rev{
\salim{Some recent works, such as~\cite{9218533} and \cite{mrazek2019autoax}, use automated search methods to provide libraries consisting of hundreds of approximate versions of an operator with varying accuracy-performance trade-offs.}
However, in most of these works, the design space exploration involves identifying feasible design points \salim{from an existing library} of 
logic-optimized operators. \salim{For example, the works presented in~\cite{9218533} and \cite{mrazek2019autoax} limit the design space to the designs presented for \glspl{asic} in~\cite{9233379}. Similarly, the works presented in~\cite{10.1145/3195970.3196115, 9344673} offer a limited design space for \gls{fpga}-optimized approximate multipliers.}
While this approach reduces the design space considerably, it may lead to reduced quality of results owing to the reduced explorable design space. 
For the extensive class of error-tolerant applications being 
\salim{explored,}
it is necessary to generate novel platform-specific approximate operators that help satisfy an application's overall accuracy-performance constraints. 
To this end, some works have provided automated exploration frameworks to generate novel 
\salim{platform-specific}
approximate implementations of arithmetic operators. 
For example, the authors of~\cite{10.1145/2966986.2967021} have used \gls{cgp} to design \gls{asic}-optimized unsigned approximate multipliers for \glspl{ann}. 
However, this work has not considered the performance metrics, such as dynamic power consumption, while generating the new designs. 
Recently, some works have reported automated \gls{dse} for 
\salim{\gls{fpga}-optimized}
\glspl{axo} using AI/ML-based methods~\cite{ullah2022appaxo,10.1145/3566097.3567891}. 
Authors of \cite{ullah2022appaxo} proposed a novel operator model for the automated synthesis of novel approximate LUT-level \salim{optimized} designs and used \gls{ga}-based search with ML-based fitness functions.
The work presented in~\cite{10.1145/3566097.3567891} has used \glspl{gan} to identify operator configurations satisfying the provided accuracy-performance constraints. 
}

\salim{\autoref{table:rel} summarizes the various aspects explored across multiple related works on the design of \gls{fpga}-based \glspl{axo} and highlights the novel contributions of our proposed work. The different approaches can be categorized into -- 1) \textit{Implementing} \gls{asic}-optimized logic designs on \glspl{fpga}~\cite{9218533}, 2) \textit{Synthesizing} a limited library of \gls{fpga}-optimized approximate operators that allows manual methods to search for a feasible operator for an application~\cite{10.1145/3195970.3196115,8465781,9344673}, and 3) \textit{Synthesizing} novel operators by optimizing for \gls{fpga}'s \gls{lut} and \gls{cc} structures and employing automated search methods to identify feasible operators~\cite{10.1145/3566097.3567891,ullah2022appaxo}. To the best of our knowledge, none of the related \gls{ml}-based \gls{dse} approaches use any information/knowledge from the characterization of smaller bit-width operators while designing larger bit-width \glspl{axo}. We posit -- \textit{the knowledge derived from statistical analysis of smaller bit-width operators can improve the quality of results in the automated \gls{dse} of larger bit-with operators.} To enable such an exploration, \titleName~presents a framework for building models that can be leveraged by metaheuristic optimization algorithms.}

\section{Operator Model}
\label{sec:op_model}
\begin{figure}[t] 
    \centering
      \scalebox{1}{\includegraphics[width=1 \columnwidth]{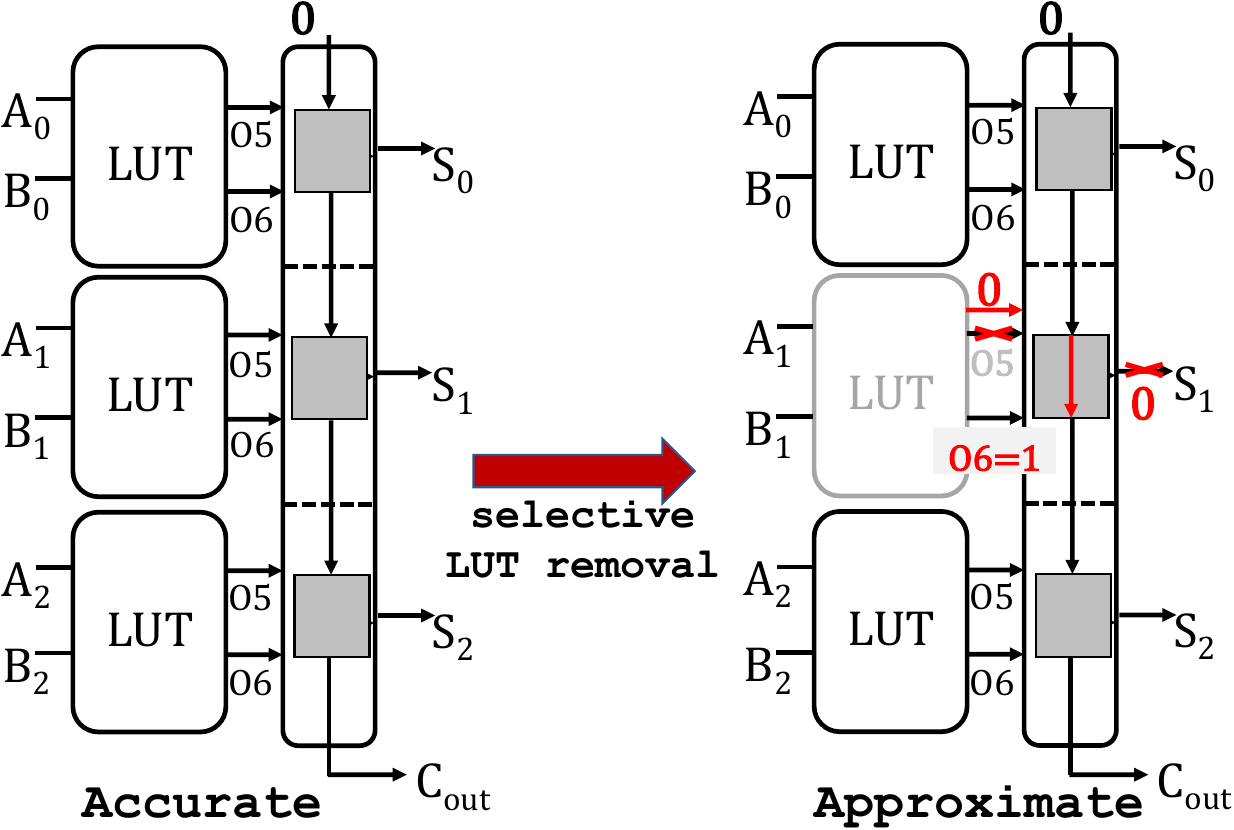}}
  \caption{Approximating a $3-bit$ unsigned adder's FPGA implementation using selective LUT removal\salim{\cite{ullah2022appaxo}}}
  \label{fig:op_model} 
\end{figure}
\rev{
For the current article, we use an operator model similar to that proposed in~\cite{ullah2022appaxo} and also used in~\cite{10.1145/3566097.3567891}. However, the proposed methods can be extended to other operator models and algorithms that can potentially exhibit similar correlations among operators of different bit widths~\cite{10.1145/3583781.3590222}. For such operator models, any \gls{fpga}-based arithmetic operator can be represented by an ordered tuple $\mathcal{O}_i(l_0,l_1,..., l_l,..., l_{L-1}), \forall l_l \in \left\{ 0,1 \right\}$. The term $l_l$ represents whether the \gls{lut} corresponding to the operator's accurate implementation is being used or not, and $L$ represents the total number of \glspl{lut} of the accurate implementation that may be removed to implement approximation. So, the accurate implementation can be represented as $\mathcal{O}_{Ac}(1,1,...,1)$. For instance, the accurate implementation of the  3-bit unsigned adder, shown in~\autoref{fig:op_model}, is represented by the tuple (1,1,1). Similarly, $\mathcal{O}=\left\{ \mathcal{O}_i \right\}$ represents the set of all possible implementations of the operator. Therefore, the set  $\mathcal{O}$ for the adder shown in the figure is \{ (0,0,0), (0,0,1), (0,1,0),(0,1,1), (1,0,0), (1,0,1), (1,1,0), (1,1,1)\}. The approximate implementation in the figure corresponds to the tuple (1,0,1).
An arbitrary operator/application's behavior can be abstracted by a function $\mathcal{S}$. So, the operator/application output for a set of inputs can be outlined as shown in~Eq.~\eqref{equ:prob_stmnt_1}. The term $Err_{\mathcal{O}_i}$ represents the error in the operator/application's behavior (\gls{behav} metric) as a result of using an approximate operator $\mathcal{O}_i$ compared to using the accurate operator $\mathcal{O}_{Ac}$. Similarly, the operator/accelerator's hardware performance (\gls{ppa} metrics) can be abstracted as a set of functions, as shown in~Eq.~\eqref{equ:prob_stmnt_2}.
}

\rev{
\begin{equation}
\label{equ:prob_stmnt_1}
\begin{split}
    Out_{\mathcal{O}_i} = \mathcal{S}(\mathcal{O}_i,Inputs);~ 
    Out_{\mathcal{O}_{Ac}} = \mathcal{S}(\mathcal{O}_{Ac},Inputs) \\
    Err_{\mathcal{O}_i} = Out_{\mathcal{O}_{Ac}} - Out_{\mathcal{O}_i}\\
\end{split}
\end{equation}
}
\rev{
\begin{equation}
\label{equ:prob_stmnt_2}
\begin{split}
    Power~Dissipation: ~~ \mathcal{W}_{\mathcal{O}_i} = \mathcal{H}_W(\mathcal{O}_i,Inputs) \\
    LUT~Utilization: ~~ \mathcal{U}_{\mathcal{O}_i} = \mathcal{H}_U(\mathcal{O}_i) \\
    Critical~Path~Delay: ~~ \mathcal{C}_{\mathcal{O}_i} = \mathcal{H}_C(\mathcal{O}_i) \\
    Power~Delay~Product: ~~PDP_{\mathcal{O}_i} = \mathcal{W}_{\mathcal{O}_i}\times \mathcal{C}_{\mathcal{O}_i} \\
    PDPLUT_{\mathcal{O}_i} = \mathcal{W}_{\mathcal{O}_i}\times \mathcal{U}_{\mathcal{O}_i} \times \mathcal{C}_{\mathcal{O}_i}
\end{split}
\end{equation}
}


\textbf{DSE Problem Statement:} 
\rev{
The constrained search problem, with \gls{behav} and/or \gls{ppa} constraints, is shown in~Eq.~\eqref{equ:prob_stmnt_const}, where $B_{MAX}$ and $P_{MAX}$ refer to the \gls{behav} (error) and \gls{ppa} metric constraints, respectively. The \gls{dse} method is used to search for feasible Pareto-optimal solutions to this problem. 
}

\rev{
\begin{equation}
\label{equ:prob_stmnt_const}
\begin{split}
    \underset{\mathcal{O}_i \in \mathcal{O}}{\text{minimize}} ({BEHAV}_{\mathcal{O}_i},{PPA}_{\mathcal{O}_i}) \\
    s.t. ~ {BEHAV}_{\mathcal{O}_i} \leq B_{MAX} ~~ and ~~ {PPA}_{\mathcal{O}_i} \leq P_{MAX}
\end{split}
\end{equation}
}
\section{\titleName~Methodology}
\label{sec:propDSE}
\begin{figure}[t]
	\centering
	\scalebox{1}{\includegraphics[width=1 \columnwidth]{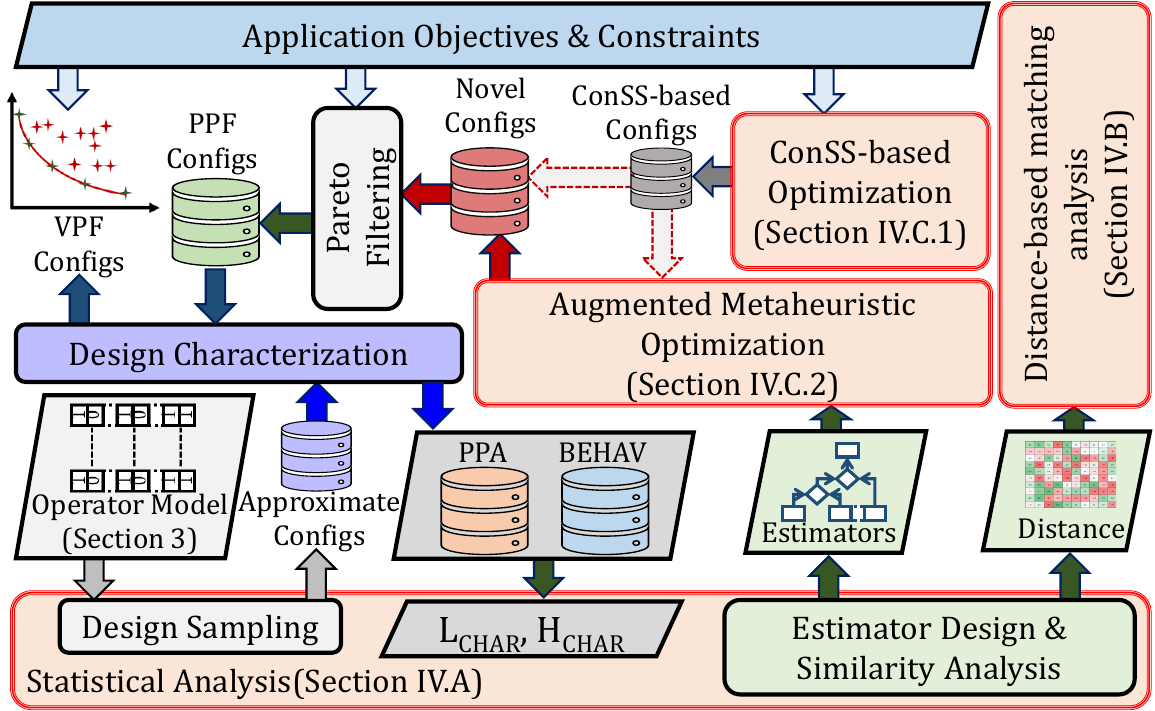}}
	\caption{\titleName~methodology}
	\label{fig:meth}
\end{figure}

\rev{
The proposed \titleName~methodology comprises three major components. 
The various processes and the corresponding flow of information amongst these components are shown in~\autoref{fig:meth}. 
\textit{Statistical Analysis} involves using the operator model for random sampling of approximate configurations. 
These configurations are then implemented and characterized for different \gls{ppa} and \gls{behav} metrics. 
$L_{CHAR}$ and $H_{CHAR}$ refer to the characterization data for low and high bit-width operators respectively. Statistical analysis also involves designing the \gls{ml}-based \textit{estimator} models and \textit{similarity analysis} across the $L_{CHAR}$ and $H_{CHAR}$ design points. The similarity analysis results are used for \textit{Distance-based Matching} to generate the modeling data for \gls{ml}-based \gls{conss}. \textit{Multiobjective \gls{dse}} involves using the \gls{ml}-based models of \gls{conss} for both standalone constrained search as well as for generating the initial solutions for an \textit{augmented metaheuristic}-based search. The \gls{ppf} solutions from the search methods are then characterized to generate the \gls{vpf} designs. The related contributions of \titleName~ are highlighted in~\autoref{fig:meth} with the corresponding section number of the article.
}

\subsection{Statistical Analysis}

\subsubsection{Estimator Design}
\rev{
Estimator design involves generating \gls{ml}-based regression models for predicting the \salim{\gls{behav}} and \gls{ppa} metrics for any arbitrary approximate configuration of an operator. It must be noted that we used \gls{ml}-based estimators only for the signed 8-bit multiplier \glspl{axo}, each represented with a 36-bit binary configuration string. For other operators, we used the metrics from the actual characterization of the \glspl{axo} as the number of possible designs, with the operator model in use, is considerably lower. Also, the current article focuses on using the characterization data more efficiently than just being limited to designing surrogate functions for estimating \gls{ppa} and \gls{behav} metrics. Therefore, we used AutoML~\cite{mljar} to explore across \gls{ml} models and their respective hyperparameters to determine the best estimator for each metric. 
}

\subsubsection{Similarity Analysis}

\begin{figure*}[t] 
    \centering
  \subfloat[PDPLUT]{
    \scalebox{1.0}{
    \includegraphics[width=0.95 \columnwidth]
        {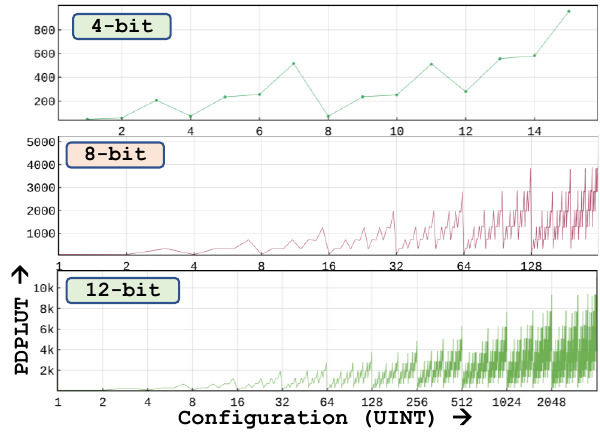}}
   }
  \subfloat[AVG\_ABS\_REL\_ERR]{
    \scalebox{1.0}{\includegraphics[width=0.95 \columnwidth]
        {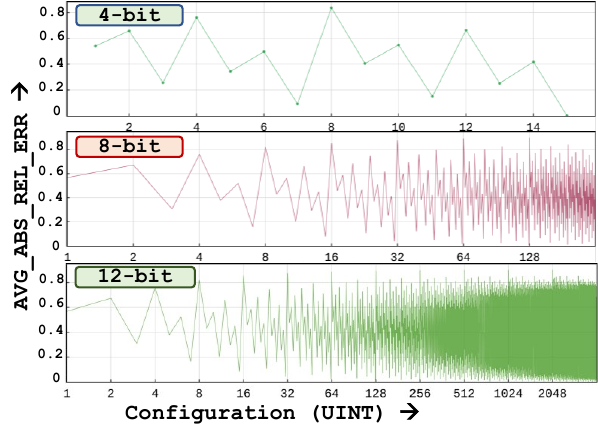}}
   }
\caption{Configuration-PPA/BEHAV trends for unsigned adder AxOs}
  \label{fig:exp_stat_trnds} 
\end{figure*}

\rev{
Similarity analysis involves the investigation into the relationship between the \gls{axo}'s configuration string and the corresponding \gls{ppa} and \gls{behav} metrics across functionally similar operators of varying operand bit-widths. While the eventual target is to create a dataset of matching approximate configurations across different bit-width operators, we begin with the analysis of the $L_{CHAR}$ and $H_{CHAR}$ datasets.
The variations of PDPLUT and AVG\_ABS\_REL\_ERR with the UINT encoding of the approximate configurations for 4-, 8-, and 12-bit unsigned adders are shown in~\autoref{fig:exp_stat_trnds}. 
\salim{The data shown in the figure is similar to that shown in ~\autoref{fig:mot_2}; however, it does not employ sub-sampling.}
The 8-bit and 12-bit results are shown with a logarithmic scale on the horizontal axis. As can be seen in the figure, the pattern of variations is similar across all the operators. This observation further motivates the analysis of the \textit{Configuration-\gls{behav}/\gls{ppa}} correlation across operators of varying bit-widths. 
\salim{It should be observed that each approximate operator can be represented by both the approximate configuration (a binary list/string) as well as the set of \gls{behav} and \gls{ppa} metrics of the resulting implementation. Therefore, both string- and metrics-based comparisons can be used for the similarity analysis. However, we limit the exploration to \gls{ppa} and \gls{behav} metrics-based comparisons for the current article.}
}

\begin{figure}[t]
	\centering
	\scalebox{1.0}{\includegraphics[width=1 \columnwidth]{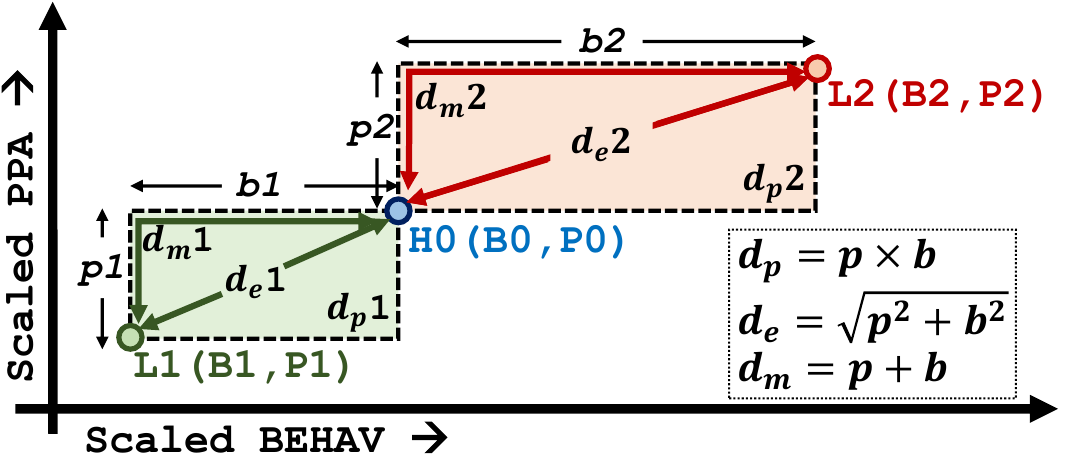}}
	\caption{Similarity Analysis}
	\label{fig:meth_sim}
\end{figure}

\rev{
\textbf{Metrics comparison:} 
In this approach, the \gls{axo} designs are analyzed based on their \gls{behav} and \gls{ppa} metrics. The k-means clustering results shown in~\autoref{fig:mot} is an example of such an analysis using Euclidean distance. However, other distance metrics can also be used for analysis. \autoref{fig:meth_sim} shows three different distance measures using the \gls{behav} \salim{and} \gls{ppa} metrics as coordinates of a Cartesian system. \salim{For example,} $H0(B0,P0)$ corresponds to a design point from the $H_{CHAR}$ dataset, with higher bit-width operands. \salim{In this representation, $B0$ and $P0$ identify the horizontal and vertical coordinates of the design point, respectively.} Similarly, $L1(B2,P2)$ and $L2(B2,P2)$ represent two designs from the $L_{CHAR}$ dataset. The corresponding distance measures are:
\begin{itemize}
    \item \textit{Euclidean distance}, $d_e$, represents the traditional measure of closeness of two points and is used in analyzing the spread of the actual \gls{ppa} and \gls{behav} metrics.
    \item \textit{Pareto distance}, $d_p$, represents a more \gls{dse}-specific measure of the similarity of two design points. It represents a more relativistic measure than $d_e$.
    \item \textit{Manhattan distance}, $d_m$, represents a similar measure as $d_p$, albeit with a slower growth rate.
\end{itemize}
While the distance measures discussed above represent unsigned values, a sign can be added to the distance values to represent the relative location of the points. For instance, in~\autoref{fig:meth_sim} points $L1$ and $L2$ can have the same distance measures from $H0$. However, adding a sign to represent whether $B0$ and/or $P0$ is lesser than $B1$/$B2$ and/or $P1$/$P2$ provides information regarding their relative location.
}

\subsection{Distance-based matching}
\begin{figure}[t]
	\centering
	\scalebox{1}{\includegraphics[width=1 \columnwidth]{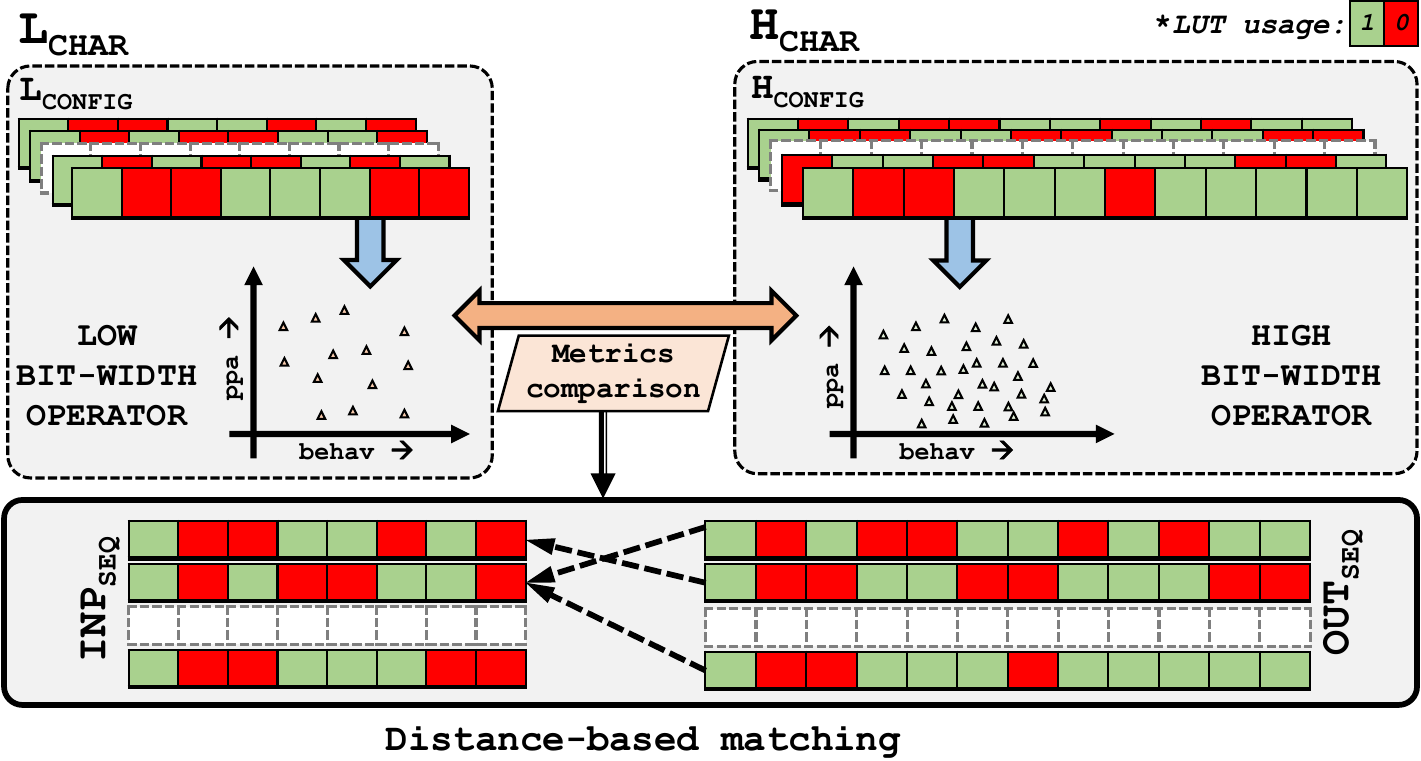}}
	\caption{Distance-based Matching}
	\label{fig:meth_dist_match}
\end{figure}

\rev{
The similarity measures presented here are not claimed to be exhaustive or the most appropriate for the current problem statement. Other methods, such as cosine distance and compression-based string matching, can also be used for the similarity analysis. Also, the perfect distance measure for the current problem can be a combination of a subset of BEHAV/PPA metrics- and string-comparison algorithms. Such an exploration is beyond the scope of the current article, and we limit our current work to show the usefulness of similarity analysis in improving the efficacy of the \gls{dse} problem by facilitating \gls{ml}-based supersampling.
}

\rev{
Distance-based matching is used to generate a dataset of input and output binary sequences, from the $L_{CHAR}$ and $H_{CHAR}$  approximate operator configurations respectively, that can be used for \gls{conss}. \autoref{fig:meth_dist_match} shows the general idea behind distance-based matching. The quantitative distance measures derived from the similarity analysis, using any of the distance measures, are used to determine the distance of each configuration in $H_{CHAR}$ to each configuration in $L_{CHAR}$. The $L_{CONFIG}$ that is least distant, say $l_{min}$, from each $H_{CONFIG}$, say $h_{i}$, is added to the resulting dataset as an $INP_{SEQ} \rightarrow OUT_{SEQ}$ pair. As shown in the figure, multiple $h_{i}$ configurations may have the same $l_{min}$ configuration leading to a \textit{one-to-many} mapping.
}

\subsection{Multiobjective DSE}
\subsubsection{\acrfull{conss}}
\begin{figure}[t]
	\centering
	\scalebox{1}{\includegraphics[width=1 \columnwidth]{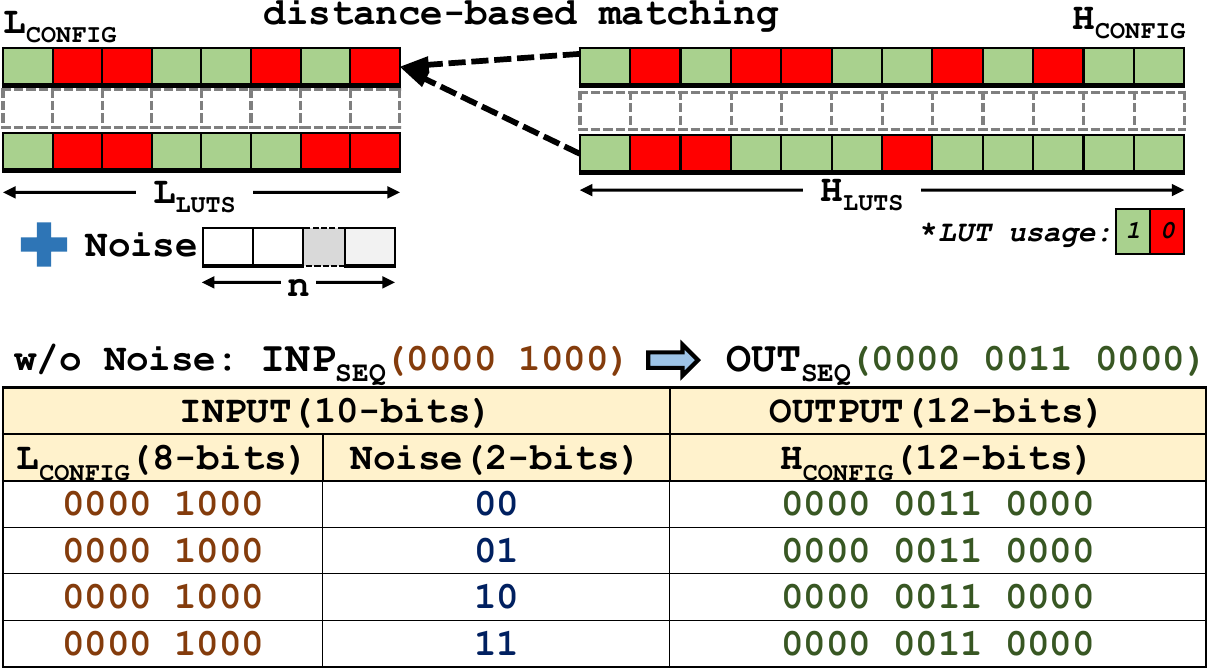}}
	\caption{Dataset generation for configuration supersampling}
	\label{fig:dse_conss}
\end{figure}
\rev{
With the distance-based matching described above, using any arbitrary similarity algorithm (or a combination of them) will result in a separate dataset of $INP_{SEQ} \rightarrow OUT_{SEQ}$ pairs. Each of these datasets can be used to train an \gls{ml} model for supersampling. For a given problem statement, such as Eq.~\eqref{equ:prob_stmnt_const}, the $L_{CONFIG}s$ satisfying the scaled constraints can be used to predict $H_{CONFIG}s$ that will also satisfy the constraints. However, as mentioned in the \aspdac{previous section}, lower bit-width operators (with smaller approximate binary configuration lengths) will have a lower number of design points in the BEHAV-PPA cartesian plane. As a result, supersampling one $L_{CONFIG}$ will result in a single $H_{CONFIG}$. However, this approach does not allow us to generate a large set of possible $H_{CONFIG}$ solutions. Furthermore, as we use supersampled solutions based on \gls{conss} to enhance a metaheuristic-based search, we would prefer to generate a large, and possibly diverse, number of possible solutions to direct the search.
}

\rev{
We propose two approaches to obtain larger and more diverse potential $H_{CONFIG}$ solutions. Firstly, we can use multiple similarity measure algorithms, each one generating its own dataset for training \gls{ml} models. Second, as shown in~\autoref{fig:dse_conss}, we add noise sequences to the $L_{CONFIG}$s. Using a noise sequence of $n$ bits allows us to generate $2^n$ $INP_{SEQ} \rightarrow OUT_{SEQ}$ pairs for each original $INP_{SEQ} \rightarrow OUT_{SEQ}$ pair. As shown in the figure, using a 2-bit noise sequence results in 4 samples for the same pair. During supersampling on trained models, theoretically, each $L_{CONFIG}$ can be used to generate 2\textsuperscript{n} $H_{CONFIG}$ potential solutions.
}

\subsubsection{Augmented Metaheuristics-based DSE}
\begin{figure}[t]
	\centering
	\scalebox{1}{\includegraphics[width=1 \columnwidth]{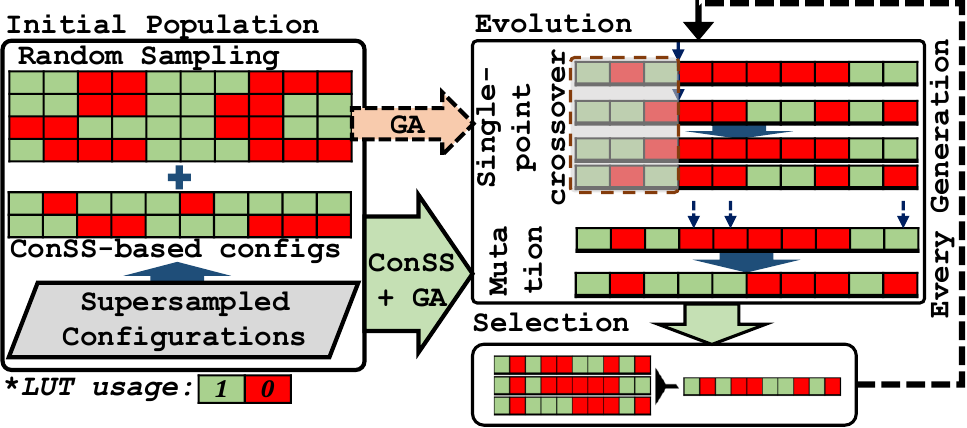}}
    \caption{Augmented GA-based DSE in~\titleName.}
	\label{fig:dse_aug_ga}
\end{figure}
\rev{
We use \acrfull{ga} as an example of a metaheuristics-based solver for the \gls{dse} of \gls{fpga}-based \glspl{axo}. 
\gls{ga} involves generating an initial population of sample solutions and selecting the population for the next generation from a set of solutions obtained by crossover and mutation of the current population. 
We used tournament selection and single-point crossover with a maximum of 250 generations for each experiment.
As shown in~\autoref{fig:dse_aug_ga}, the problem-agnostic \gls{ga} method involves using random sampling to generate the initial population. We also use an augmented approach where we use the solution pool generated by \gls{conss} as the initial population, in addition to the random initial configurations. 
This allows us to direct the search toward Pareto-optimal solutions faster using supersampling.
}
\section{Experiments and Results}
\label{sec:expRes}

\begin{table}[htb!]
\centering
\caption{
Integer Arithmetic Operators Used for \\Experimental Evaluation of \titleName 
}
\label{table:conss}
\def\arraystretch{1.2}
\resizebox{1 \columnwidth}{!}{
\begin{tabular}{|c|c|c|c|cll|}
\hline
\begin{tabular}[c]{@{}c@{}}Operator \\ Type\end{tabular} & \begin{tabular}[c]{@{}c@{}}Operand\\ Bit-width\end{tabular} & \begin{tabular}[c]{@{}c@{}}Possible AxO\\ Designs\end{tabular} & \begin{tabular}[c]{@{}c@{}}Configuration\\ String Length\end{tabular} & \multicolumn{3}{c|}{\begin{tabular}[c]{@{}c@{}}ConSS\\ upscale factor\end{tabular}} \\ \hline
\multirow{3}{*}{\begin{tabular}[c]{@{}c@{}}Unsigned \\ Adder\end{tabular}} & 4-bits & 16 & 4-bits & \multicolumn{1}{c|}{\multirow{2}{*}{\textdownarrow2x}} & \multicolumn{1}{l|}{} & \multirow{3}{*}{\textdownarrow3x} \\ \cline{2-4} \cline{6-6}
 & 8-bits & 255 & 8-bits & \multicolumn{1}{c|}{} & \multicolumn{1}{l|}{\multirow{2}{*}{\textdownarrow1.5x}} &  \\ \cline{2-5}
 & 12-bits & 4096 & 12-bits & \multicolumn{1}{c|}{} & \multicolumn{1}{l|}{} &  \\ \hline
\multirow{2}{*}{\begin{tabular}[c]{@{}c@{}}Signed \\ Multiplier\end{tabular}} & 4-bits & 1024 & 10-bits & \multicolumn{3}{c|}{\multirow{2}{*}{\textdownarrow3.6x}} \\ \cline{2-4}
 & 8-bits & 68.7 Billion & 36-bits & \multicolumn{3}{c|}{} \\ \hline
\end{tabular}
}
\end{table}
\subsection{Experiment Setup}

\rev{
\autoref{table:conss} shows the operator designs used in the experiments for evaluating the proposed methods in \titleName.
For the multipliers, we used the Baugh-Wooley algorithm-based implementation as the baseline accurate design.
While random sampling was used for generating the approximate configurations for the signed $8\times8$ multiplier designs, all possible \gls{axo} designs were characterized for the other arithmetic operators in the table.
\salim{As highlighted in~\autoref{sec:op_model}, the proposed methodology can be effectively employed for diverse arithmetic algorithms, such as Booth's multiplication algorithm. Furthermore, it can also be utilized for other operator models that exhibit analogous correlations between operators with different bit widths.}
\autoref{table:conss}~also shows the scale-up factor, based on the configuration lengths, when using \gls{conss} across functionally similar operators of different bit-width operands.
These configurations are implemented in VHDL and synthesized for the $7VX330T$ device of the Virtex-7 family using Xilinx Vivado 19.2. 
The synthesis and implementation of each configuration involved multiple executions where we updated the critical path constraint according to the previously achieved critical path slack to obtain highly precise \gls{cpd} and dynamic power consumption values for each design. 
The \gls{ml}-based supersampling and \gls{dse} methods are implemented in Python, utilizing packages such as DEAP and Scikit among others.
}


\subsection{Statistical Analysis \& Distance-based Matching}
\label{subsec:expres_stat_dist}

\begin{figure}[t]
    \centering
  \subfloat[Absolute metrics]{
    \scalebox{1}{\includegraphics[width=0.34 \columnwidth]{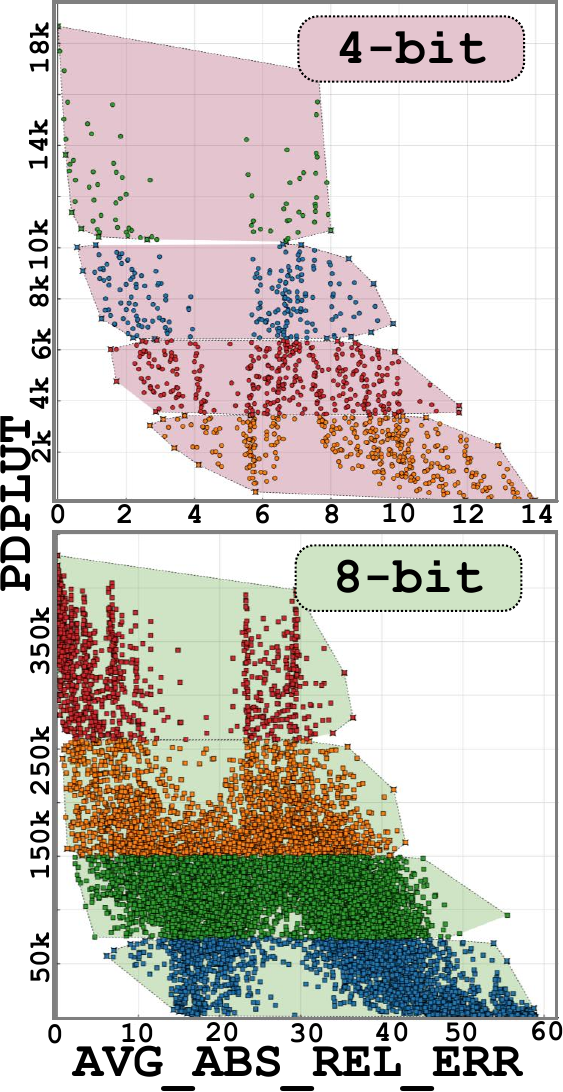}}
       }
  \subfloat[Scaled metrics]{%
        \scalebox{1}{\includegraphics[width=0.66 \columnwidth]{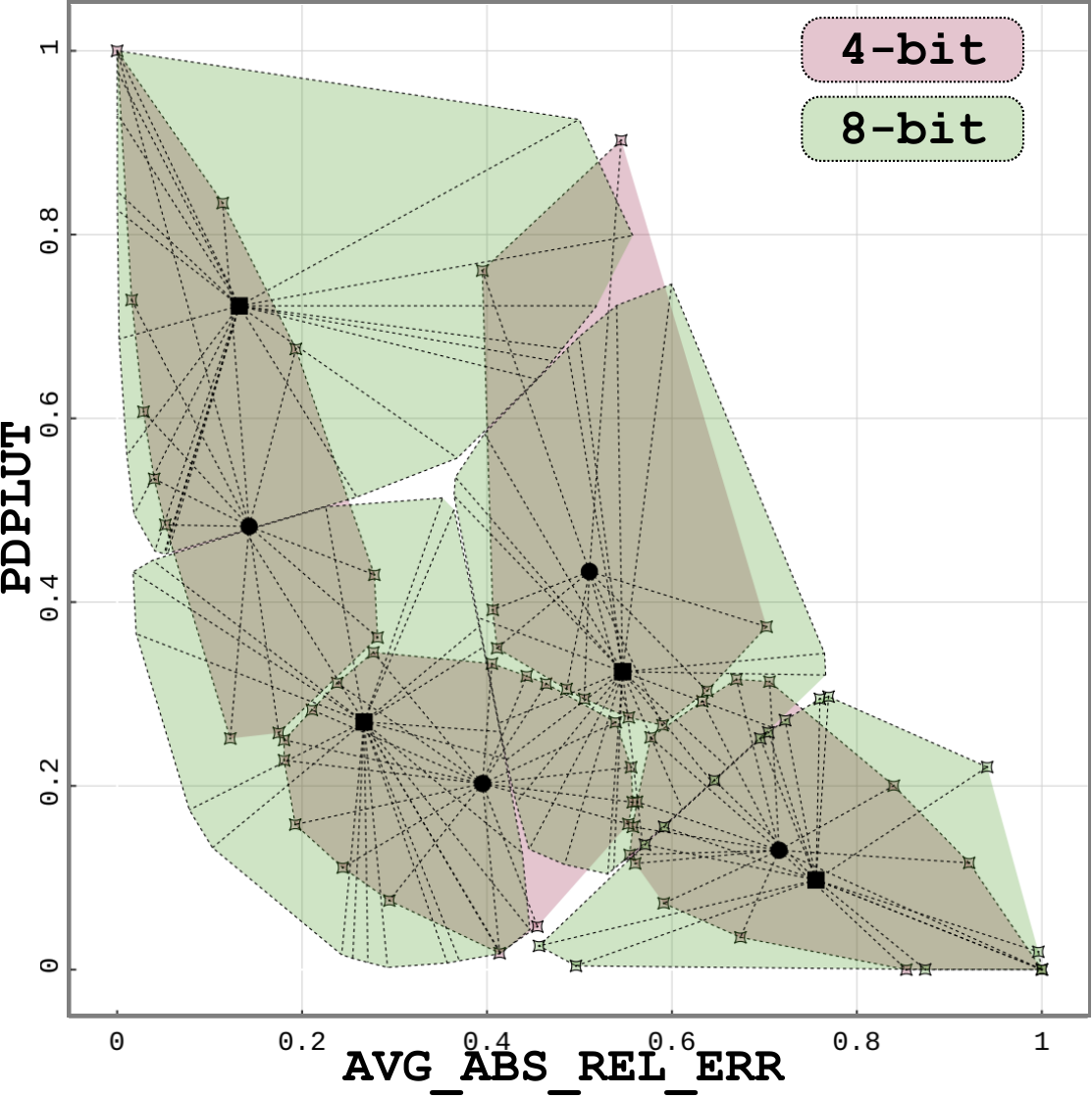}}
        }
\caption{k-means clustering of designs points representing approximate implementations of 4-bit and 8-bit signed multipliers}
  \label{fig:exp_stat_cluster} 
\end{figure}

\rev{
As described earlier, we used AutoML for designing the \gls{ml}-based estimators of PPA and BEHAV metrics of 8-bit signed approximate multipliers. Since the features of the characterization data are essentially categorical variables (1/0, denoting if a LUT is being used or not), ML models based on CatBoost and LightGBM exhibited the best trade-offs of accuracy and R2 score for the training and testing datasets. Metrics that are products of other metrics such as PDP and PDPLUT reported larger  \gls{rmse} values than individual metrics. For the analysis of the patterns of design points of varying operator widths, we employed a k-means clustering similar to~\autoref{fig:mot}. The results of the k-means clustering for the 4-bit and 8-bit signed multiplier \glspl{axo} are shown in \autoref{fig:exp_stat_cluster}. Although the analysis shows an equal number of clusters for both operator sizes, the centroids for both cases do not align as well as that for the adders. This can be attributed to the \textit{non-exhaustive} samples for the 8-bit signed multiplier \glspl{axo}. While the 4-bit approximate multiplier dataset has all the possible 1023 \footnote{We do not use the configuration with all 0s in the experiments.}design points, only 10,650, out of a possible 68.7 billion, points are sampled and characterized for the 8-bit approximate multipliers.
}

\begin{figure}[t] 
	\centering
	\scalebox{1}{\includegraphics[width=1 \columnwidth]{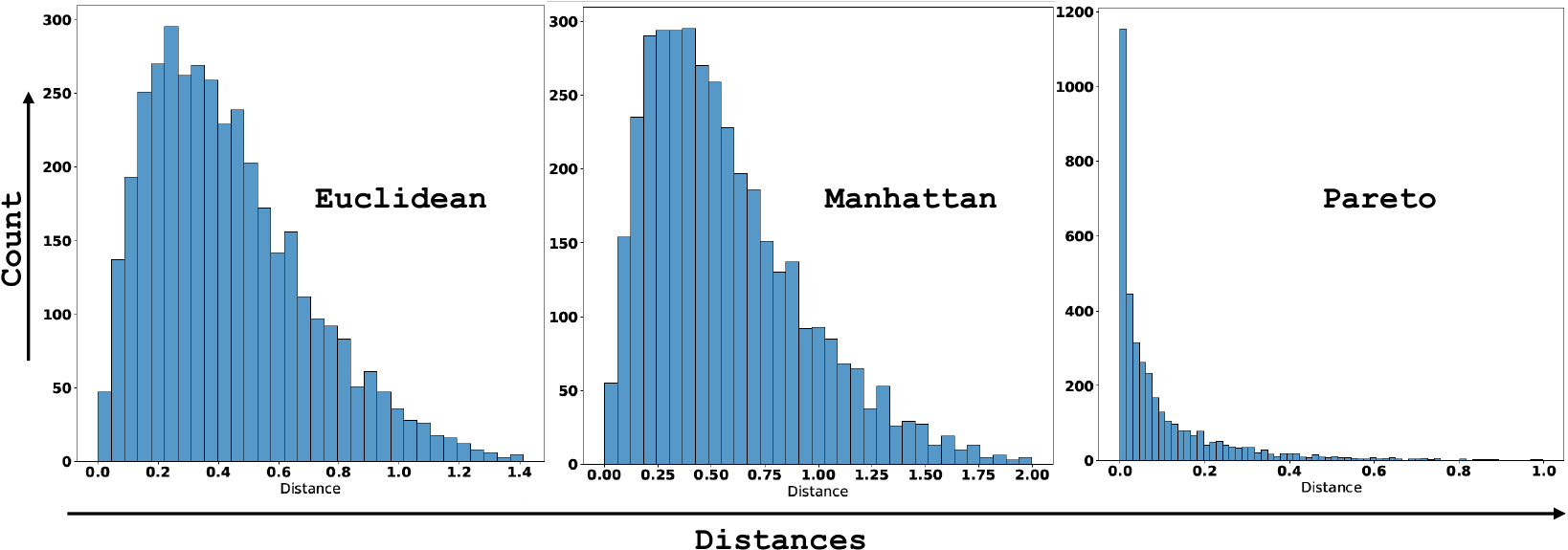}}
	\caption{Distribution of distances for unsigned 4- and 8-bit AxO adders}
	\label{fig:exp_stat_hist_uadd}
 \vspace{-10pt}
\end{figure}

\rev{Another aspect of the statistical analysis involves computing the distance of each low-bit-width operator from each high-bit-width operator. 
\autoref{fig:exp_stat_hist_uadd} shows the distribution of such distances among the unsigned 4- and 8-bit approximate adders. 
\salim{The figure shows that}
the Euclidean and the Manhattan distance measures show similar distributions with a considerable spread of values. However, the Pareto distance measure shows a more long-tailed distribution with few values showing up as the distance between 
\salim{many}
 operator pairs. A wider distribution, \aspdac{as exhibited by Euclidean- and Manhattan-based measures,} indicates a better differentiation of the pairs of \glspl{axo} as possible matching configurations. Similar distributions were also observed for signed \salim{$4\times4$ and $8\times8$} approximate multipliers.
}

\begin{figure}[t]
    \centering
  \subfloat[Euclidean Distance]{
    \scalebox{1.0}{
    \includegraphics[width=0.52 \columnwidth]{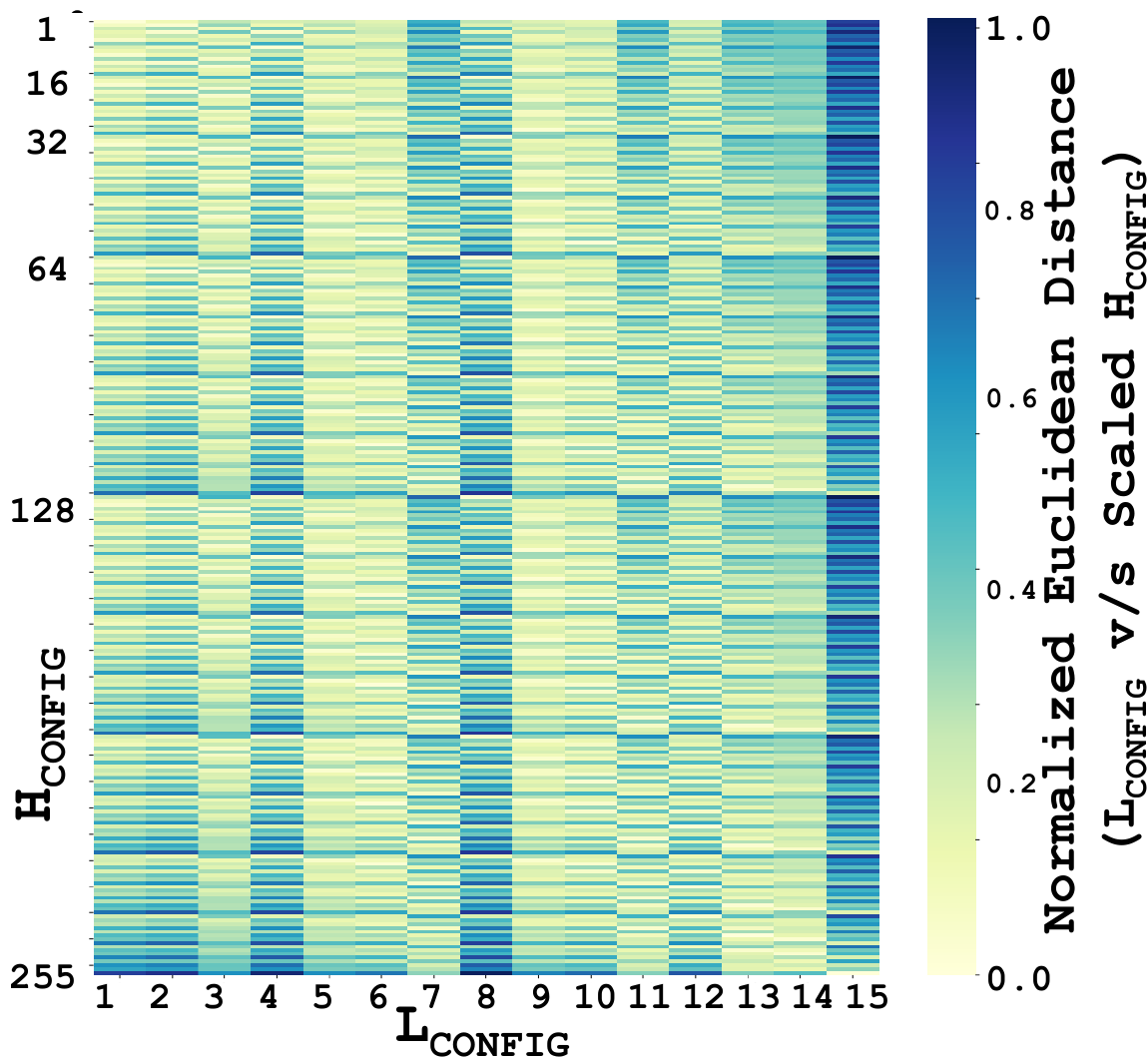}}
       }
  \subfloat[Matched Configurations]{%
        \scalebox{1.0}{\includegraphics[width=0.48 \columnwidth]{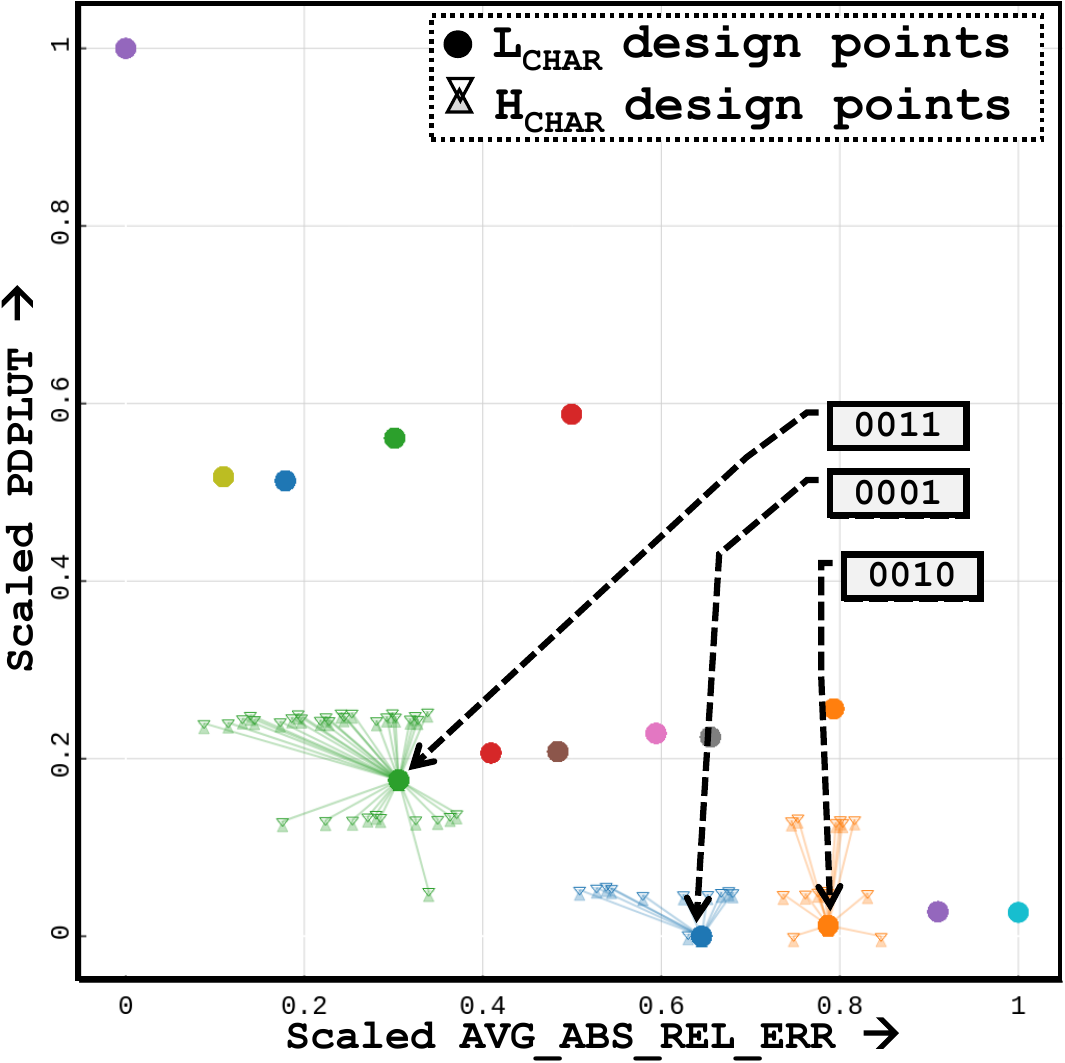}}
        }
\caption{Distance-based matching for 4-bit and 8-bit unsigned adder AxOs}
  \label{fig:meth_dist_match_res} 
\end{figure}

\rev{
\salim{We utilized Euclidean distance-based matching to generate the training dataset for the \gls{ml}-based supersampling.}
\autoref{fig:meth_dist_match_res} shows the results of a sample analysis between \glspl{axo} for unsigned 4-bit ($L_{CHAR}$) and 8-bit ($H_{CHAR}$) adders. \autoref{fig:meth_dist_match_res}(a) shows the heat-map of the Euclidean distances between $L_{CONFIG}$ and $H_{CONFIG}$ configurations. \autoref{fig:meth_dist_match_res}(b) shows the corresponding distance matching of 3 (out of 15)
unique configurations of the 4-bit unsigned adder \glspl{axo} along with all the 8-bit designs matched with each of them. As shown in \autoref{fig:meth_dist_match_res}(b), 33, 12, and  14 different $H_{CONFIG}$s map into the $L_{CONFIG}s$ 0011, 0001, and 0010 respectively.
}

\subsection{\acrfull{conss}}


\begin{figure}[htb!]
	\centering
	\scalebox{1}{\includegraphics[width=1 \columnwidth]{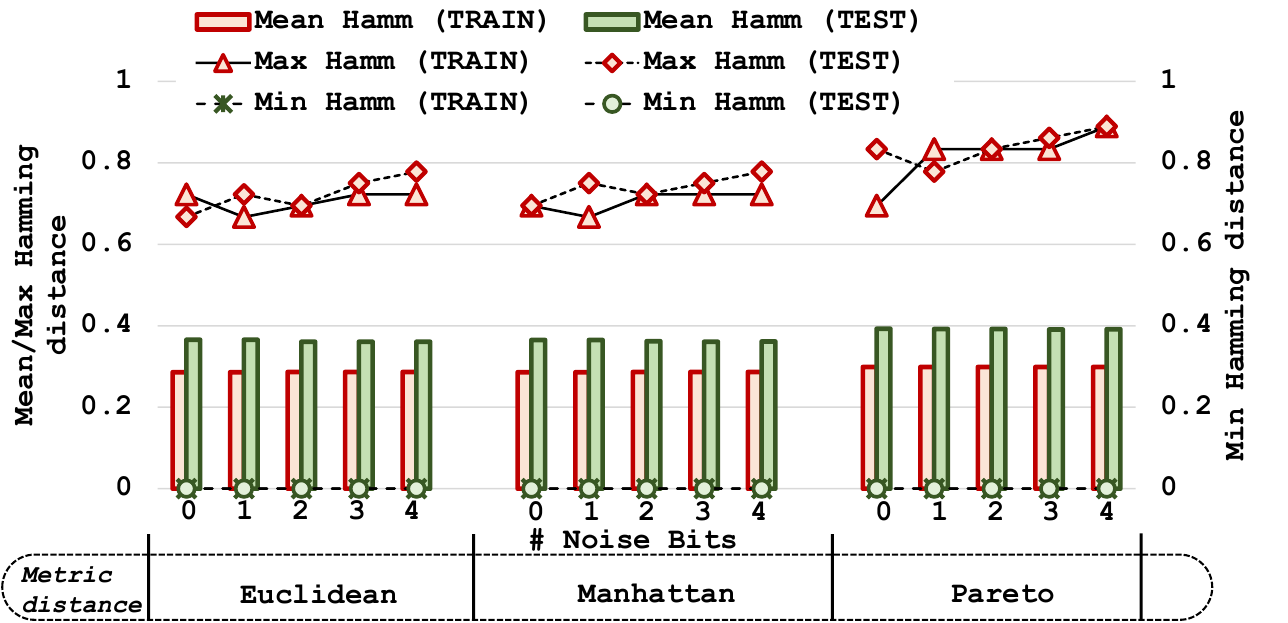}}
	\caption{ML Modelling for ConSS using Random Forest-based Multi-output Classification}
	\label{fig:exp_conss_model_ml}
\end{figure}

\begin{figure}[htb!]
	\centering
	\scalebox{1}{\includegraphics[width=1 \columnwidth]{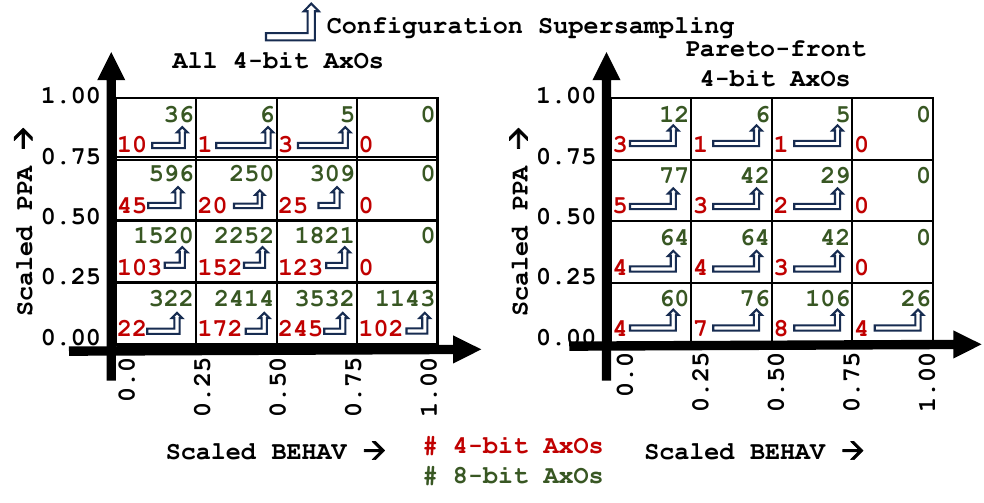}}
	\caption{ConSS of signed \salim{$4\times4$} approximate multipliers}
	\label{fig:exp_conss_grid}
\end{figure}

\rev{
\aspdac{Based on the distribution analysis, better spread, and the similarity with the Manhattan distance measure, we used the Euclidean distance as the metric used in supersampling. The corresponding distance-matched datasets were used for ML modeling for \gls{conss}. }\autoref{fig:exp_conss_model_ml} shows the results of using Random Forest-based multi-output classification for the \salim{$4\times4$ to $8\times8$} signed multiplier \gls{conss}. This entails predicting a 36-bit binary string from a 10-bit binary string corresponding to the approximate configurations of the \salim{$8\times8$ and $4\times4$} signed multiplier \glspl{axo}, respectively. We quantify the resulting prediction in terms of the Hamming distance between the original and predicted output sequences. \autoref{fig:exp_conss_model_ml} also shows the variation of the accuracy with an increasing number of noise bits. As observed in the figure, using additional noise bits does not affect the accuracy of the model.
}

\rev{
\autoref{fig:exp_conss_grid} shows the number of design points for the $4\times4$ \aspdac{\glspl{axo} (in green)} and the number of unique design points predicted for the $8\times8$ \glspl{axo} in different regions of the \gls{behav}-\gls{ppa} plane. While the left figure uses all the designs in each region for \gls{conss}, the figure on the right uses only the Pareto-front design points in each region for \gls{conss}. The \gls{conss} was achieved by varying the number of noise bits in the \gls{conss} \gls{ml} models trained with Euclidean distance-matching.
}

\subsection{Multi-objective Design Optimization}

\begin{figure}[t]
	\centering
	\scalebox{1}{\includegraphics[width=1 \columnwidth]{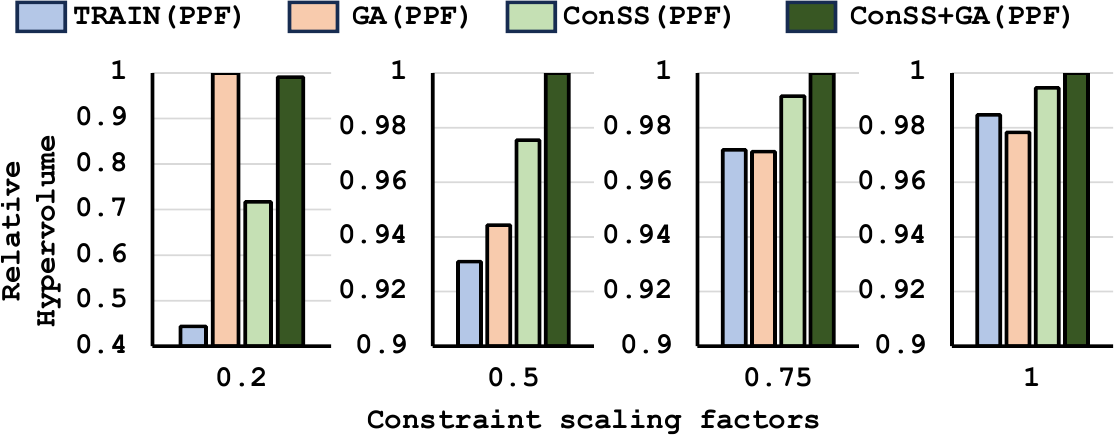}}
	\caption{ConSS+GA  DSE results for 8-bit signed approximate multipliers}
	\label{fig:exp_conss_ga_dse}
\end{figure}

\rev{
The experiments for \titleName-based \gls{dse} involved comparing the resulting Pareto-front hypervolume of the designs generated with \gls{conss} both as a standalone method as well as augmenting \gls{ga}-based exploration for signed $8\times8$ approximate multipliers. Hypervolume is a well-studied method for assessing multiobjective optimization and is estimated as the area (for two objectives) swept by a point or Pareto-front w.r.t. a reference point, usually defined by the problem's constraints. We used a \textit{constraint scaling factor} to evaluate the results under varying levels of constrained optimization.
The scaling factor corresponds to the value that is multiplied by the maximum value of \gls{ppa} and \gls{behav} in the training dataset of 10,650 points to obtain the $P_{MAX}$ and $B_{MAX}$ of Eq.~\eqref{equ:prob_stmnt_const} respectively. A smaller constraint scaling factor implies a more tightly constrained problem. 
Standalone \gls{conss} results in considerably improved hypervolume over the training data (up to 40\%), specifically for tightly constrained problems. The \gls{vpf} results are obtained using the characterization of 31, 282, 365, and 390 additional \gls{axo} configurations for scaling factors of 0.2, 0.5,0.75, and 1.0 respectively.
}


\begin{figure}[t]
	\centering
	\scalebox{1}{\includegraphics[width=1 \columnwidth]{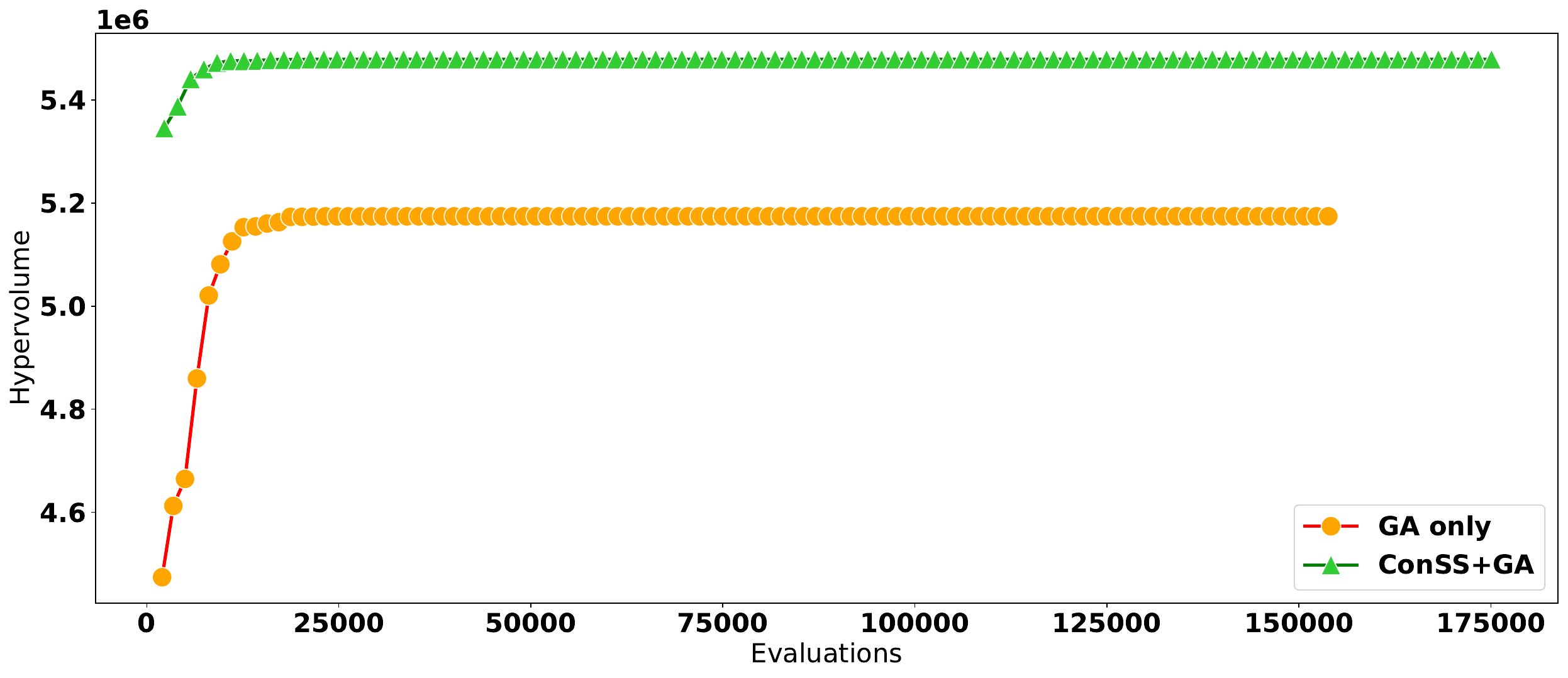}}
	\caption{Progress of hypervolume with a constraint scaling factor of 0.5}
	\label{fig:exp_dist_ga_progress}
\end{figure}

\rev{
The design points obtained with \gls{conss} were used as initial solutions for a \gls{ga}-based exploration. \autoref{fig:exp_conss_ga_dse} shows the final Pareto-front hypervolume obtained from the training data, \gls{ga}-only, \gls{conss}, and \gls{conss}+GA. The comparisons correspond to varying constraint scaling factors and are obtained from the \gls{ppf}, based on the predicted \gls{ppa} and \gls{behav} metrics using the \gls{ml}-based estimators during \gls{ga} evolution. As seen in the figure, \gls{conss}+\gls{ga} results in improvements over the non-augmented \gls{ga}. The progression of the hypervolume during \gls{ga}-based evolution for both cases is shown in \autoref{fig:exp_dist_ga_progress}. As evident, the \gls{conss}+\gls{ga} approach starts with much better solutions, owing to the results from \gls{conss} and ends with far better hypervolume.
}

\subsection{Comparing with State-of-the-art}
\begin{figure}[t]
\vspace{5pt}
	\centering
	\scalebox{1}{\includegraphics[width=1 \columnwidth]
    {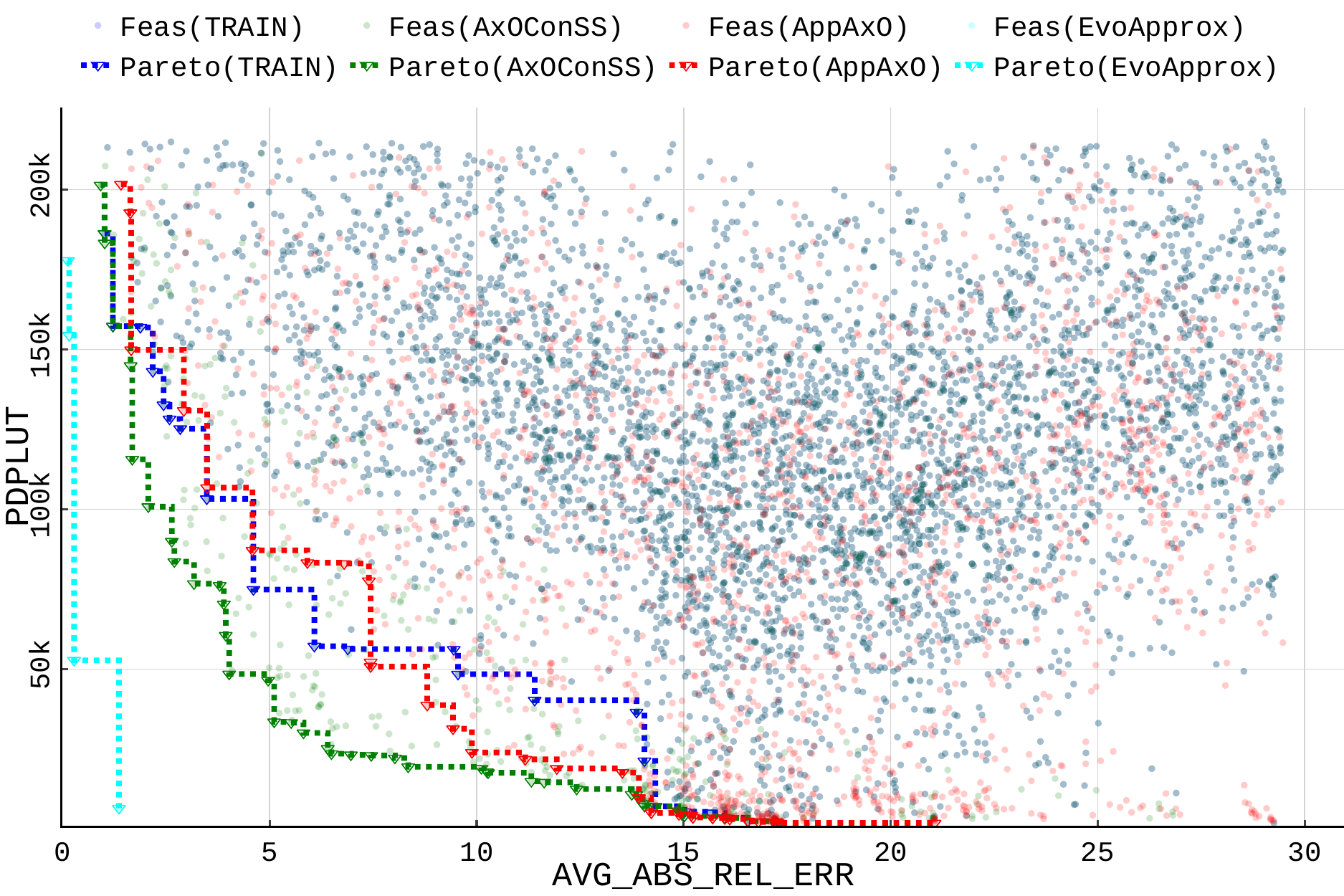}}
    \caption{Pareto fronts for 8-bit signed approximate multipliers}
	\label{fig:exp_soa}
\end{figure}

\begin{figure}[t]
	\centering
	\scalebox{1}{\includegraphics[width=1.0 \columnwidth]{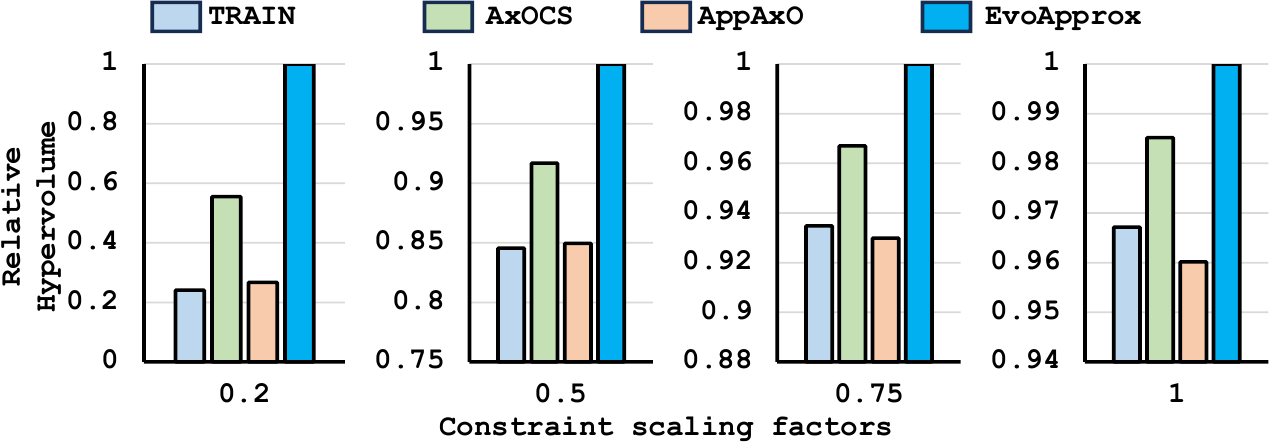}}
	\caption{Comparing Relative Hypervolume}
	\label{fig:exp_soa_all}
\end{figure}

\rev{
\autoref{fig:exp_soa} shows the \gls{vpf} obtained with \titleName~compared to the training data (TRAIN), AppAxO~\cite{ullah2022appaxo}, and EvoApprox~\cite{9233379} based 8-bit signed approximate multipliers\footnote{We do not compare our results with \cite{10.1145/3566097.3567891,10.1145/3583781.3590222} as the resulting designs from those works are not available.}. The plot shows the results for a constraint scaling factor of 0.5. The comparison of the hypervolume at all the scaling factors is shown in \autoref{fig:exp_soa_all}. As can be seen, \titleName~ reports considerably better results than AppAxO and similar results at EvoApprox for loosely constrained problems. As was mentioned in AppAxO, the limitations of the operator model limit the quality of results compared to EvoApprox. Improved models can be used with the proposed \titleName~ methodology for better quality results, especially for application-specific design.
}


\section{Conclusion}
\label{sec:conc}


\rev{
The current article proposes a novel approach for the DSE of FPGA-based approximate arithmetic operators. The proposed framework uses supersampling of design configurations to improve the efficacy of traditional optimization methods and leverages the more elaborate exploration possible at lower bit-width operators to scale up the operators. 
Related future research could involve evaluating the proposed approach for application-specific design and using other \textit{distance} comparison methods along with deploying \textit{sequence-to-sequence} models for supersampling.
}     

\balance
\bibliographystyle{IEEEtran}
\bibliography{references}
\newpage
\begin{IEEEbiography}[{\includegraphics[width=1in,height=1.25in,clip]{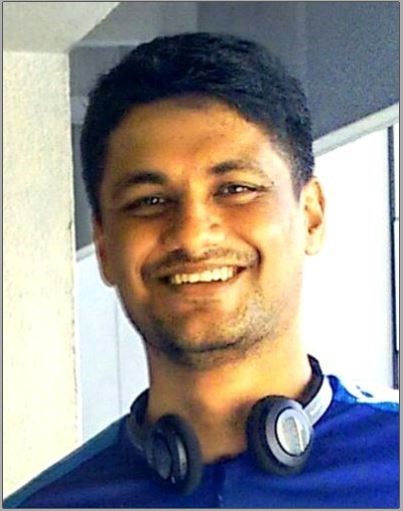}}]{Siva Satyendra Sahoo} is currently working as a R\&D Engineer (Specialist) with IMEC, Leuven.  He received his doctoral degree (Ph.D., 2015-2019) in the field of reliability in heterogeneous embedded systems from the National University of Singapore, Singapore. He completed his masters (M.Tech, 2010-2012) from the Indian Institute of Science, Bangalore in the specialization Electronics Design Technology. He has also worked with Intel India, Bangalore in the domain of Physical Design. His research interests include Embedded
Systems, Machine Learning, Approximate Computing, Reconfigurable Computing, Reliability-aware Computing Systems, and System-level Design. For the presented work, he was working as Postdoctoral Researcher with the Chair for Processor Design at TU Dresden.
\end{IEEEbiography}

\begin{IEEEbiography}[{\includegraphics[width=1in,height=1.25in,clip]{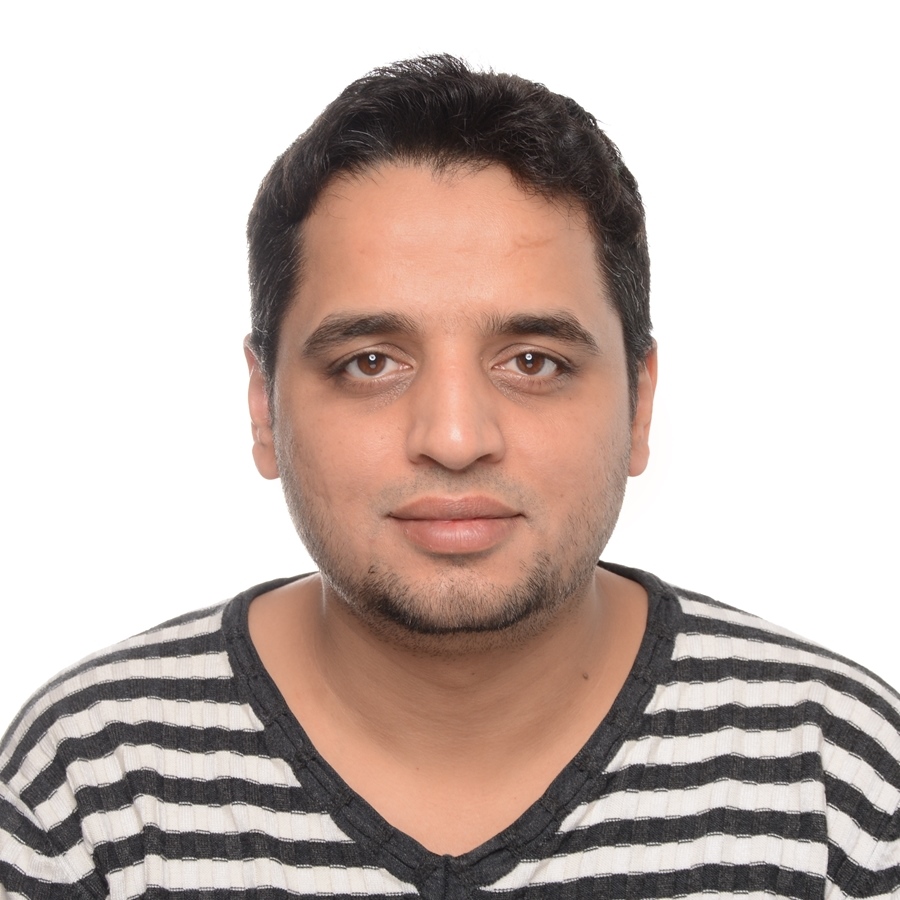}}]{Salim Ullah} is a Postdoctoral Researcher at the Chair for Processor Design, Technische Universit{\"a}t Dresden. He has completed his BSc and MSc in Computer Systems Engineering from the University of Engineering and Technology Peshawar, Pakistan. His current research interests include the Design of Approximate Arithmetic Units, Approximate Caches, and Hardware Accelerators for Deep Neural Networks. 
\end{IEEEbiography}

\begin{IEEEbiographynophoto}{Soumyo Bhattacharjee} is a Masters student at the Eidgenössische Technische Hochschule Zürich. He has completed his B.tech in Electrical and Electronics Engineering from the Indian Institute of Technology, Kharagpur, India. For the presented work, he was working for his research internship with the Chair for Processor Design at TU Dresden.
\end{IEEEbiographynophoto}

\begin{IEEEbiography}[{\includegraphics[width=0.95in,height=1.25in,clip]{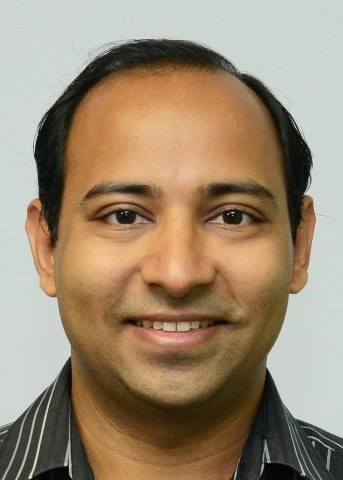}}]{Akash Kumar} (SM’13) received the joint Ph.D. degree in electrical engineering and embedded systems from the Eindhoven University of Technology, Eindhoven, The Netherlands, and the National University of Singapore (NUS), Singapore, in 2009. From 2009 to 2015, he was with NUS. He is currently a Professor with Technische Universität Dresden, Dresden, Germany, where he is directing the Chair for Processor Design. His current research interests include the design, analysis, and resource management of low-power and fault-tolerant embedded multiprocessor systems.
\end{IEEEbiography}
\end{document}